\documentclass[a4paper,fleqn]{cas-sc}

\usepackage[square, comma, sort&compress, numbers]{natbib}

\usepackage{amsmath}
\usepackage[linesnumbered,boxed]{algorithm2e}
\usepackage[justification=centering]{caption}
\usepackage{caption}
\usepackage{subcaption}
\usepackage{graphicx}
\usepackage{longtable}
\usepackage{multirow}
\usepackage{array}
\usepackage{pdflscape}
\usepackage{booktabs}
\usepackage[figuresright]{rotating}
\usepackage{multicol}
\newproof{pf}{Proof}
\usepackage{lineno}
\usepackage{float}
\def\tsc#1{\csdef{#1}{\textsc{\lowercase{#1}}\xspace}}
\tsc{WGM}
\tsc{QE}
\tsc{EP}
\tsc{PMS}
\tsc{BEC}
\tsc{DE}

\begin{document}
\let\printorcid\relax
\let\WriteBookmarks\relax
\def\floatpagepagefraction{1}
\def\textpagefraction{.001}

\title [mode = title]{Adaptive Vehicle Speed Classification via BMCNN with Reinforcement Learning-Enhanced Acoustic Processing}

\author[1]{Yuli Zhang\textsuperscript{\ddag}}[]
\ead{Yuli.Zhang20@student.xjtlu.edu.cn}



\author[1]{Pengfei Fan\textsuperscript{\ddag}}  
\ead{Pengfei.Fan22@student.xjtlu.edu.cn} 
\author[1]{Ruiyuan Jiang}  
\ead{Ruiyuan.Jiang20@student.xjtlu.edu.cn} 
\author[1]{Hankang Gu}  
\ead{Hankang.Gu16@student.xjtlu.edu.cn} 
\author[1]{Dongyao Jia*}  
\ead{Dongyao.Jia@xjtlu.edu.cn} 
%
\author[1]{Xinheng Wang*}  
\ead{Xinheng.Wang@xjtlu.edu.cn} 
\cortext[cor1]{Corresponding authors: Dongyao Jia, Xinheng Wang \\  \hspace*{3ex}  \textsuperscript{\ddag}These authors contributed equally to this work.  }

\affiliation[1]{organization={School of Advanced Technology},
	addressline={Xi'an Jiaotong-Liverpool University}, 
	city={Suzhou},
	postcode={215123}, 
	country={China}}

\newcommand{\figref}[1]{Fig.~\ref{#1}}

\begin{abstract}
Traffic congestion persists as a pressing challenge in contemporary urban environments, demanding sophisticated intelligent transportation systems (ITS) for effective real-time traffic management and infrastructure optimization. While conventional sensor-based methods incur substantial installation and maintenance expenses, acoustic sensors offer a cost-efficient alternative for traffic surveillance. This study introduces an innovative hybrid methodology integrating deep learning with reinforcement learning for enhanced acoustic-based vehicle speed classification. We develop a Bidirectional Multi-modal Convolutional Neural Network (BMCNN) architecture that processes dual acoustic representations through specialized CNN branches, employing MFCC and wavelet features to effectively capture vehicle acoustic patterns across different frequency domains. To overcome computational efficiency limitations in real-time implementations, we propose a Deep Q-Network with attention mechanism (AttentionDQN) that adaptively determines the optimal number of audio frames required for precise classification, facilitates early decision-making when confidence thresholds are met. The proposed system undergoes rigorous evaluation using both the publicly accessible IDMT-Traffic dataset and our newly collected SZUR-Acoustic dataset from urban environments in Suzhou, China. Experimental findings indicate that our BMCNN-AttentionDQN framework achieves 95.99\% accuracy on IDMT-Traffic and 92.3\% on SZUR-Acoustic datasets, while achieving up to 1.63× reduction in average processing time through intelligent early termination strategies. The framework demonstrates superior performance compared to state-of-the-art baselines including A3C, DDDQN, SA2C, PPO, and TD3, effectively resolving the inherent accuracy-efficiency trade-off. This research advances acoustic-based traffic monitoring by delivering an adaptive, efficient, and accurate vehicle speed classification solution suitable for real-time ITS deployment in heterogeneous urban settings.
\end{abstract}

\begin{highlights}
	\item Novel BMCNN architecture with dual-stream acoustic processing combining MFCC and wavelet features for robust vehicle speed               classification
	\item Attention-DQN with adaptive early stopping mechanism reduces processing time by up to 1.63× while maintaining high accuracy
	\item Theoretical proof of Markov property establishes rigorous foundation for applying reinforcement learning to acoustic processing efficiency optimization
	\item Comprehensive evaluation on two datasets achieves 95.99\% accuracy on IDMT-Traffic and 92.3\% on SZUR-Acoustic, outperforming five state-of-the-art baselines
\end{highlights}





\maketitle

\section{Introduction}
Traffic congestion is a severe problem worldwide. Due to the increase in motor vehicles, accelerated urbanization, and population growth, global traffic congestion has become increasingly serious. Traffic congestion reduces the efficiency of urban transportation infrastructure, increases travel time, fuel consumption, and air pollution, and also leads to dissatisfaction among and fatigue of urban residents. Accurate vehicle speed estimation and classification play a crucial role in intelligent transportation systems (ITS), enabling real-time traffic management, congestion control, and infrastructure optimization \cite{tyagi2012vehicular, djukanovic2021acoustic}.

An increasing number of countries have proposed various intelligent transportation system (ITS) solutions based on lane-driving facts and suitable for routine deployment. These solutions rely on multiple sensors that are typically expensive, including magnetic loop sensors, speed detectors, and video cameras. Currently, magnetic loop detectors are the most commonly used traffic sensors in developed countries \cite{papageorgiou2003review}. However, the installation and maintenance costs of these sensors significantly increase operational expenses throughout their lifecycle. The associated costs of road construction and traffic control further limit their widespread adoption. Consequently, throughout the past decade, researchers have been developing various non-intrusive traffic monitoring technologies based on laser, ultrasonic, radar, video, and audio signals.

Video image processing appears to be a natural sensing approach for traffic monitoring. Despite remarkable progress in recent years, state-of-the-art computer vision techniques remain insufficient for detecting heterogeneous objects under various lighting conditions and occlusion scenarios—a critical requirement for any video-based traffic monitoring system. For instance, Kamijo et al. \cite{kamijo2000traffic} presented a Hidden Markov Model (HMM)-based computer vision technique for detecting accidents and other events such as reckless driving at intersections. However, these techniques do not directly address the problem of vehicle average speed/speed range estimation and suffer from environmental sensitivity.

In contrast, acoustic sensors have emerged as a cost-effective alternative, providing extensive road-related data including vehicle noise data, vehicle counting, and vehicle estimation. These sensors offer numerous practical applications in traffic monitoring systems, such as automatic incident detection through abnormal sound pattern recognition, traffic density estimation in tunnels and underpasses where visual monitoring is challenging, and weather condition assessment by analyzing tire-road interaction sounds. Acoustic sensors have proven effective for classifying different vehicle types based on their distinct acoustic signatures \cite{cevher2009vehicle}, enabling more granular traffic analysis for urban planning and congestion management. Additionally, these sensors can operate continuously in all weather conditions and lighting scenarios, making them particularly suitable for 24/7 traffic monitoring in harsh environments where traditional camera-based systems may fail.

Recent advances in deep learning and signal processing have further enhanced the capabilities of acoustic-based traffic monitoring systems. Machine learning algorithms, particularly convolutional neural networks (CNNs), have demonstrated superior performance in extracting meaningful features from acoustic signals for traffic parameter estimation \cite{salamon2017deep, george2020vehicle}. Acoustic road monitoring systems have shown remarkable success in detecting anomalous sounds and estimating traffic parameters with high accuracy \cite{foggia2016audio}. Deep learning approaches have revolutionized acoustic traffic monitoring by enabling automatic feature extraction from raw audio signals, eliminating the need for manual feature engineering that characterized traditional signal processing methods. These advances have enabled the development of real-time monitoring systems with accuracy comparable to traditional sensor networks while maintaining significantly lower infrastructure costs and complexity.

Despite these advances, current acoustic-based vehicle speed classification systems face several critical limitations. First, existing deep learning models often require processing complete audio sequences to achieve optimal accuracy, leading to increased computational overhead and delayed response times in real-time applications \cite{lan2022dynamic}. Second, most current approaches lack adaptive decision-making capabilities that could optimize the trade-off between classification accuracy and computational efficiency based on real-time confidence levels. Third, early exiting mechanisms, while explored in computer vision tasks \cite{teerapittayanon2016branchynet}, have not been effectively adapted for acoustic-based traffic monitoring applications.

To address these limitations, this paper presents a novel approach for vehicle speed classification that combines advanced deep learning architectures with reinforcement learning-based optimization. The main contributions of this work are twofold:

First, we propose a Bidirectional Multi-modal Convolutional Neural Network (BMCNN) architecture specifically designed for acoustic-based vehicle speed classification. The BMCNN model processes dual acoustic representations (MFCC and wavelet features) through specialized CNN branches, enabling robust feature extraction from vehicle acoustic signatures while maintaining computational efficiency. We validate the effectiveness of our BMCNN approach on both the publicly available IDMT-Traffic dataset \cite{abesser2021idmt} and a comprehensive SZUR-Acoustic dataset collected through our own field measurements, demonstrating superior performance compared to existing state-of-the-art methods.

Second, we introduce a Deep Q-Network with attention mechanism (AttentionDQN) that dynamically determines the optimal number of audio frames required for accurate classification. Unlike traditional approaches that process fixed-length audio sequences, our RL-enhanced system learns to make early classification decisions when confidence levels are sufficiently high, significantly reducing computational overhead without compromising accuracy. The AttentionDQN incorporates adaptive reward mechanisms and attention-based state representations to optimize the accuracy-efficiency trade-off in real-time applications.

The proposed hybrid approach addresses the fundamental challenge in acoustic-based ITS: achieving high classification accuracy while maintaining real-time processing capabilities. By combining the feature extraction strength of BMCNN with the adaptive decision-making capabilities of AttentionDQN, our system achieves remarkable performance improvements. Experimental results demonstrate that our method achieves classification accuracy of 95.99\% on the IDMT-Traffic dataset and 92.3\% on the SZUR-Acoustic dataset, while reducing average processing time by up to 1.63$\times$ through intelligent early stopping mechanisms. The attention-enhanced DQN framework effectively balances the accuracy-efficiency trade-off, outperforming state-of-the-art baselines including A3C, DDDQN, SA2C, PPO, and TD3.

The remainder of this paper is organized as follows: Section \ref{S2} presents the related work in acoustic traffic monitoring and reinforcement learning applications. Section \ref{S3} formulates the problem statement. Section \ref{S4} describes the proposed BMCNN architecture and the AttentionDQN algorithm in detail. Section \ref{S5} presents the experimental setup, datasets, and evaluation metrics. Section \ref{S6} discusses the experimental results and performance comparisons. Section \ref{S7} concludes the paper and outlines future research directions.

\section{Related Work} \label{S2}
\subsection{Acoustic Traffic Monitoring Technology}
The application of acoustic sensing in traffic monitoring has advanced through successive methodological phases, evolving from threshold-based detection to conventional signal processing and, more recently, to data-driven learning paradigms. Early investigations relied on simple amplitude thresholds to determine vehicular presence, identifying pass-by events from fluctuations in ambient sound-pressure levels. With the maturation of signal-processing theory, time–frequency analysis became prevalent: spectral signatures produced by engines, tyre–road interaction, and ancillary mechanical sources were exploited for vehicle classification \cite{tyagi2012vehicular}. Short-time Fourier transforms, wavelet transforms, and power-spectral-density estimators were among the principal tools employed for detection and speed inference \cite{wu2020hdspeed,djukanovic2021acoustic,ullah2025vehicle}. In complex traffic environments these deterministic techniques exhibit limited interference immunity and modest classification accuracy, prompting the integration of statistical learning frameworks such as support-vector machines and Gaussian mixture models.

Deep learning has subsequently reshaped acoustic analysis. Convolutional neural networks automatically extract hierarchical features from Mel-spectrogram representations, markedly enhancing vehicle-type discrimination \cite{mohine2022acoustic,shams2024acoustic}. Recurrent architectures—including long short-term memory networks and their bidirectional variants—capture temporal dependencies inherent in acoustic sequences, thereby improving event-detection fidelity \cite{nithya2024tb}. Attention mechanisms direct computational focus toward salient time–frequency regions, while Transformer architectures leverage self-attention to model long acoustic sequences with parallel efficiency \cite{wang2025making}. Furthermore, multimodal fusion and end-to-end optimization unify feature extraction and decision making within a single trainable framework, reducing the error accumulation characteristic of traditional cascaded pipelines \cite{wang2022end}.

Vehicular speed estimation is currently pursued via vision-, radar-, and acoustics-based approaches. Vision techniques infer speed from inter-frame displacement yet are vulnerable to adverse illumination and meteorological conditions \cite{limit1,limit2}. Acoustic strategies deploy Doppler analysis or inter-sensor propagation-time differentials; auxiliary methods such as cross-correlation and cross-power spectral density estimation are also common. Despite their potential, acoustic schemes still face a pronounced trade-off between real-time performance and estimation accuracy: classical spectral analyses are computationally economical yet imprecise, whereas learning-based algorithms deliver superior accuracy at the expense of extensive data requirements and processing overhead \cite{djukanovic2021acoustic}. Sensitivity to environmental noise and restricted classification robustness further constrain large-scale adoption, underscoring the need for more efficient and noise-tolerant acoustic signal-processing algorithms for traffic monitoring.

\subsection{Reinforcement Learning-Driven Real-Time Traffic System Optimization}
Current reinforcement learning (RL) implementations in intelligent transportation systems face significant computational bottlenecks that limit their practical deployment in real-time environments. Traditional Proximal Policy Optimization algorithms, while demonstrating superior stability compared to other policy gradient methods, suffer from fixed-length sequence processing requirements that create unnecessary computational overhead in time-sensitive applications \cite{RL1,RL2}. Existing PPO implementations in traffic signal control achieve notable performance improvements, with systems reporting up to 27\% reduction in vehicle waiting times \cite{RL3}, yet these approaches fail to address the fundamental trade-off between classification accuracy and processing efficiency that constrains real-world deployment scenarios.

The literature reveals a critical gap in adaptive decision-making capabilities within current RL-based traffic monitoring systems. Standard PPO algorithms process complete input sequences regardless of confidence levels, leading to substantial computational waste when early classification decisions could achieve equivalent accuracy \cite{RL4,RL5}. Research in traffic signal optimization demonstrates that conventional PPO implementations require processing entire observation windows, resulting in average response times that exceed acceptable thresholds for real-time vehicle classification tasks \cite{RL6}. Furthermore, existing approaches lack dynamic adaptation mechanisms that could optimize computational resource allocation based on varying environmental conditions and system performance requirements.

Recent advances in acoustic-based vehicle monitoring have highlighted the potential for reinforcement learning optimization, yet current methodologies remain constrained by computational efficiency limitations \cite{RL7,RL8}. Traditional CNN-based approaches for acoustic vehicle classification achieve accuracies ranging from 86\% to 98\% but require fixed computational budgets that prevent adaptive optimization \cite{RL9,RL10}. The integration of reinforcement learning with acoustic processing systems represents an underexplored domain where adaptive learning mechanisms could substantially improve both accuracy and efficiency performance metrics.

The proposed improved Attention-DQN algorithm addresses these fundamental limitations through several key innovations that establish clear advantages over existing methodologies. First, the algorithm incorporates confidence-based early termination mechanisms that enable dynamic sequence length determination, reducing average processing time by up to 1.63× while maintaining comparable accuracy levels to full-sequence processing approaches. This adaptive decision-making capability represents a significant advancement over static processing frameworks that characterize current implementations \cite{RL11,RL12}.

Second, the improved Attention-DQN algorithm integrates attention-based state representations that enhance feature extraction efficiency while reducing computational complexity. Unlike conventional approaches that process audio frames uniformly, the proposed method dynamically allocates computational resources based on signal importance, achieving superior performance with substantially reduced processing overhead \cite{RL13}. The attention mechanism enables the algorithm to focus on acoustically significant portions of vehicle sound signatures, eliminating redundant processing of low-information audio segments.

Third, the algorithm implements adaptive reward mechanisms that optimize the accuracy-efficiency trade-off in real-time based on system performance feedback. This dynamic optimization capability enables the system to adjust processing intensity according to current computational constraints and accuracy requirements, providing flexibility that existing static optimization approaches cannot achieve \cite{RL14,RL15}. The adaptive reward structure ensures optimal performance across varying operational conditions while maintaining system reliability and responsiveness.

The integration of these innovations within the hybrid BMCNN-AttentionDQN framework creates synergistic effects that exceed the performance capabilities of individual optimization approaches. The bi-directional multi-scale CNN architecture provides robust feature extraction capabilities, while the improved Attention-DQN algorithm optimizes the computational efficiency of the classification process through intelligent early stopping and adaptive resource allocation \cite{RL16}. This combination addresses the critical challenge in acoustic-based intelligent transportation systems of achieving high accuracy while maintaining real-time processing capabilities.

\section{Problem Statement}\label{S3}

The acoustic-based vehicle speed classification problem inherently involves balancing classification accuracy against computational efficiency—a challenge that existing approaches have yet to adequately resolve. Given continuous acoustic signals captured by roadside sensors, the task is to classify vehicle speeds into discrete categories (low, medium, high) while maintaining minimal processing latency for real-time deployment.

Formally, let $\mathbf{X} = \{x_1, x_2, ..., x_T\}$ denote a sequence of acoustic frames, where $x_t \in \mathbb{R}^d$ represents the $d$-dimensional feature vector at time $t$, and $T$ is the maximum sequence length. The classification function $f: \mathbf{X} \rightarrow \mathcal{Y}$ maps acoustic sequences to speed categories $\mathcal{Y} = \{y_{\text{low}}, y_{\text{mid}}, y_{\text{high}}\}$. Conventional approaches process the entire sequence $\mathbf{X}$ to produce classification decisions, incurring a fixed computational cost of $\mathcal{O}(T)$ regardless of sample complexity or confidence levels.

The core challenge involves identifying the minimal subsequence $\mathbf{X}_{1:t^*} = \{x_1, ..., x_{t^*}\}$ where $t^* \leq T$ that contains sufficient information for reliable classification. This necessitates an intelligent agent capable of evaluating current classification confidence and making real-time stop/continue decisions. Consequently, we must solve the following joint optimization problem:

\begin{equation}
\min_{t^*, \theta} \mathcal{L}_{\text{acc}}(f_\theta(\mathbf{X}_{1:t^*}), y) + \lambda \cdot \mathcal{C}(t^*)
\label{eq:optimization}
\end{equation}

\noindent where $\mathcal{L}_{\text{acc}}$ denotes the classification loss, $\mathcal{C}(t^*)$ represents the computational cost proportional to $t^*$, $\lambda$ controls the accuracy-efficiency trade-off, and $\theta$ encompasses the model parameters.

This optimization faces several key challenges:

\begin{enumerate}
    \item The optimal stopping point $t^*$ varies across samples based on acoustic complexity and environmental conditions
    \item Stopping decisions must be made without future frame information, necessitating real-time decision-making
    \item The system must demonstrate robustness across diverse acoustic environments while adhering to strict real-time constraints
\end{enumerate}

Our proposed BMCNN-AttentionDQN framework addresses these challenges through adaptive processing that dynamically balances accuracy and efficiency according to each input sequence's unique characteristics.

\section{Methodology}\label{S4}

\subsection{Feature Extraction}
\subsubsection{Mel-frequency Cepstral Coefficients}
In speech and audio research, Mel-frequency cepstral coefficients (MFCCs) remain a staple representation. The usual pipeline can be summarized as follows: the raw waveform is firstly conditioned by noise suppression, short-time framing, and windowing to ensure local stationarity and mitigate edge discontinuities; each windowed frame is then converted to the frequency domain through a fast Fourier transform (FFT); the resulting magnitude spectrum passes through a bank of triangular filters distributed on the Mel scale, thereby emulating the ear's non-linear frequency selectivity; the energies at the filter outputs are log-compressed to curb dynamic-range variation; finally, the relationship that warps a physical frequency $f$ (Hz) to its perceptual counterpart $\mu$ (Mels) is typically expressed as \cite{MFCC,MFCCM}:
\begin{equation}  
\mu = 2595 \cdot \log_{10}\left(1 + \frac{f}{700}\right) 
\end{equation}

\subsubsection{Wavelet Transform}
Wavelet analysis represents a signal by means of basis functions that are jointly localised in time and frequency, making it well suited to non-stationary data \cite{Wavelet1,Wavelet2}. Within this framework, the first-order Coiflet (Coif1) is often adopted for discrete multiresolution analysis because it combines compact support, near symmetry, and several vanishing moments. A signal $x(t)$ is successively convolved with a low-pass filter $h(n)$ and a high-pass filter $g(n)$, producing approximation and detail coefficients at multiple scales. The corresponding scaling (father) function $\phi(t)$ and mother wavelet $\psi(t)$ satisfy separate two-scale relations:

\begin{equation}
\phi(t)=\sum_{n=0}^{N-1} h(n)\phi(2t-n)
\label{eq:scaling}
\end{equation}

\begin{equation}
\psi(t)=\sum_{n=0}^{N-1} g(n)\phi(2t-n)
\label{eq:wavelet}
\end{equation}

\noindent where 
$N$ is the filter length. The scaling function captures the low-frequency, smooth content, while the wavelet isolates transient, high-frequency details—properties that underpin widespread applications of Coif1 in compression, denoising, and feature extraction.

\subsection{Frequency-Chosen Two-feature Network Architecture}
\begin{figure*}[ht]
	\centering
	\includegraphics[width=0.9\textwidth]{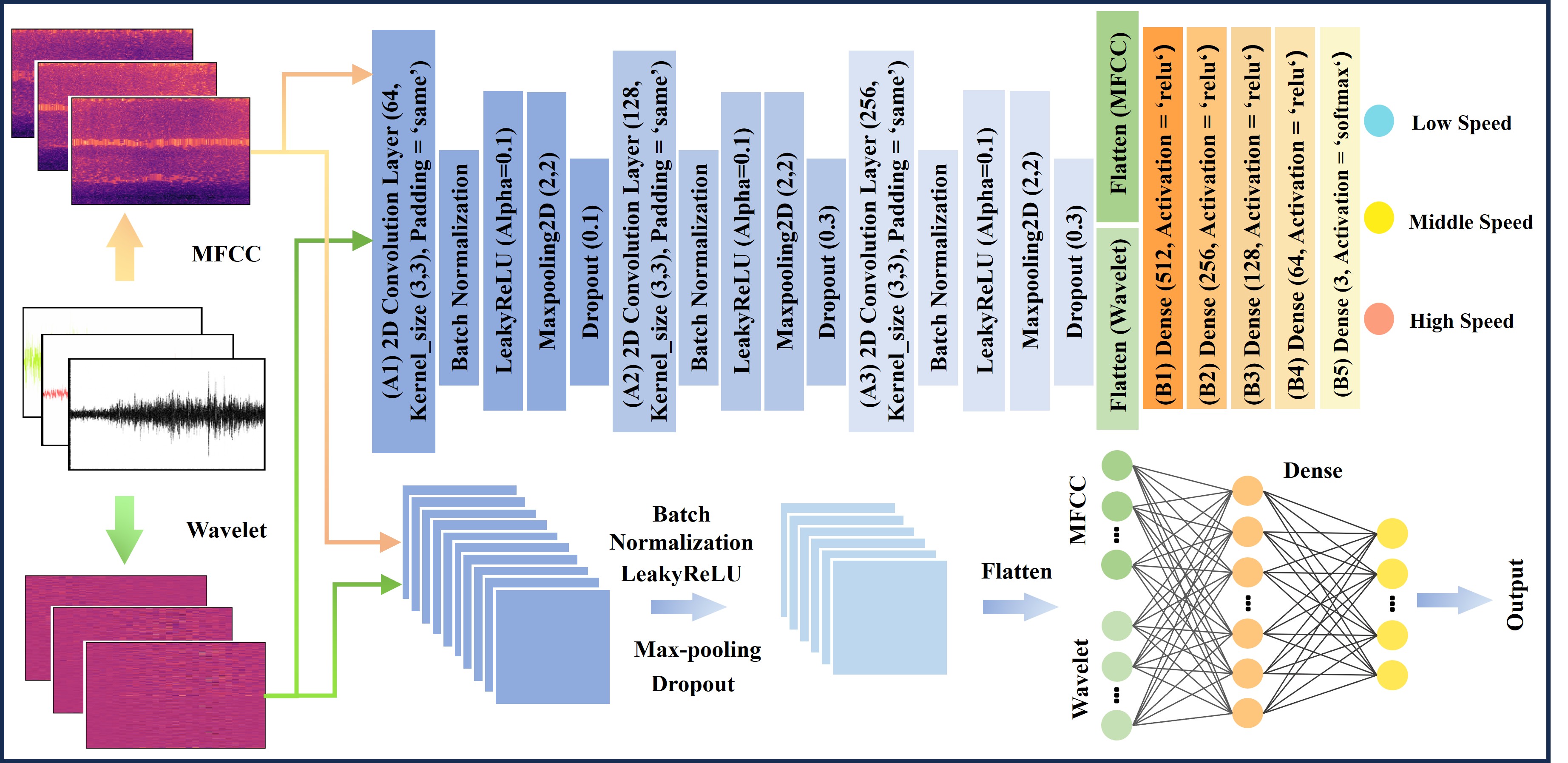}     
        \caption{Principle of Frequency-Chosen Two-feature Network Architecture.}
	\label{FS4}
\end{figure*}

Figure~\ref{FS4} presents the full workflow adopted in this study. We extract two complementary acoustic representations: (i) a Mel-frequency cepstral coefficient map $\mathbf{X}_{\mathrm{MFCC}}\in\mathbb{R}^{T\times F}$ that emphasises perceptual spectral content, and (ii) a Wavelet–PCA map $\mathbf{X}_{\mathrm{WT}}\in\mathbb{R}^{T\times F}$ that captures fine-grained time–frequency details. Treating each map as a pseudo–image allows us to process them with parallel two–dimensional convolutional neural networks. Each branch contains four convolution–pooling blocks that gradually abstract the raw features into a compact, high–level representation.

For the $\ell$–th block ($\ell=1,\ldots,4$) the forward operation is
\begin{equation}
    H_{\ell} = \mathrm{Drop} \bigl( \mathrm{MP} \bigl(\sigma (\mathrm{BN}(\mathbf{W}_{\ell} * H_{\ell-1})) \bigr) \bigr), \quad H_0 \in \{ \mathbf{X}_{\mathrm{MFCC}}, \mathbf{X}_{\mathrm{WT}} \}
\end{equation}
\noindent
where $\mathbf{W}_{\ell}$ is a learnable $k\times k$ kernel tensor, $*$ denotes 2–D convolution, $\mathrm{BN}(\cdot)$ is batch normalisation with scale–shift parameters $(\gamma, \beta)$, $\sigma(u) = \max(0, u)$ is the ReLU non–linearity, $\mathrm{MP}(\cdot)$ is $2\times2$ max–pooling, and $\mathrm{Drop}(\cdot)$ applies dropout with keeping probability $p_d$ to curb over–fitting.

After the fourth block, the resulting feature maps are flattened into vectors
\[
    \mathbf{z}_{\mathrm{MFCC}}\in\mathbb{R}^{d_1}, \qquad \mathbf{z}_{\mathrm{WT}}\in\mathbb{R}^{d_2}.
\]
These two representations are concatenated to build a fused feature vector
\begin{equation}
    \mathbf{z} = \left[ \mathbf{z}_{\mathrm{MFCC}} \, \| \, \mathbf{z}_{\mathrm{WT}} \right] \in \mathbb{R}^{d_1+d_2}.
\end{equation}

\noindent The vector $\mathbf{z}$ is then passed through two fully connected (Dense) layers:
\begin{equation}
    \mathbf{h} = \sigma(\mathbf{W}_1 \mathbf{z} + \mathbf{b}_1 ),\qquad \mathbf{o} = \mathbf{W}_2 \mathbf{h} + \mathbf{b}_2,
\end{equation}
\noindent where $\mathbf{W}_1$, $\mathbf{W}_2$ and $\mathbf{b}_1$, $\mathbf{b}_2$ are trainable weight matrices and bias vectors, and $\sigma$ is again ReLU.
The output logits $\mathbf{o}\in\mathbb{R}^3$ are converted to class probabilities for the three speed categories $\mathcal{C} = \{\mathrm{low}, \mathrm{mid}, \mathrm{high}\}$ with the Softmax function
\begin{equation}
    P(c\,|\,\mathbf{o}) = \frac{\exp(o_c)}{\sum_{c'\in\mathcal{C}} \exp(o_{c'})}, \quad c\in\mathcal{C}.
\end{equation}

\noindent Model parameters are learned by minimising the categorical cross–entropy
\begin{equation}
    \mathcal{L} = -\sum_{c\in\mathcal{C}} y_c \log P(c\,|\,\mathbf{o}),
\end{equation}

\noindent where $\mathbf{y}$ is the one–hot ground–truth label. In essence, the network first distils task–relevant cues from MFCC and Wavelet descriptors in separate convolutional hierarchies, then unifies this complementary information through feature fusion and dense layers, and finally delivers a probabilistic estimate of the engine speed class via Softmax. This design leverages both spectral and time–frequency insights, allowing the system to achieve robust and accurate audio classification without sacrificing descriptive detail.

\subsection{Markov Game Framework and Deep Q-Network with Attention}

The acoustic-based vehicle speed classification problem can be formulated as a sequential decision-making task within the Markov Game framework. In this context, the classification agent must determine the optimal stopping point for audio sequence processing while balancing accuracy requirements against computational efficiency constraints.

We model the decision process as a Markov Decision Process (MDP) defined by the tuple $\langle \mathcal{S}, \mathcal{A}, \mathcal{P}, \mathcal{R}, \gamma \rangle$, where $\mathcal{S}$ represents the state space encompassing audio feature representations and classification confidence metrics, $\mathcal{A}$ denotes the action space consisting of binary decisions to continue processing or terminate classification, $\mathcal{P}$ describes the state transition probabilities, $\mathcal{R}$ defines the reward function, and $\gamma$ represents the discount factor for future rewards.

The state representation $s_t \in \mathcal{S}$ at timestep $t$ incorporates multiple information sources derived from the neural network's predictive behavior and temporal processing characteristics. Based on our implementation, the enhanced state vector is constructed as:

\begin{equation}
s_t = \left[\frac{t}{T_{max}}, \mu_{conf}(t), \frac{H(t)}{\log C}, \hat{y}_{t,1}, \hat{y}_{t,2}, \hat{y}_{t,3}, \Delta\mu_{conf}(t), \Delta H(t), \bar{\mu}_{conf}(t), \sigma_{conf}(t), \rho_{consec}(t), \psi_{stab}(t)\right]^{\mathrm{T}}
\end{equation}

\noindent where $\frac{t}{T_{max}}$ represents the normalized temporal progress with $T_{max}$ being the maximum allowable timesteps, $\mu_{conf}(t) = \max_i \hat{y}_{t,i}$ denotes the maximum confidence across all classes, $H(t) = -\sum_{i=1}^C \hat{y}_{t,i} \log \hat{y}_{t,i}$ represents the Shannon entropy normalized by $\log C$ to quantify prediction uncertainty, $\hat{y}_{t,i}$ are the individual class probability outputs from the softmax layer, $\Delta\mu_{conf}(t) = \mu_{conf}(t) - \mu_{conf}(t-1)$ captures confidence dynamics, $\Delta H(t) = H(t) - H(t-1)$ measures entropy evolution, $\bar{\mu}_{conf}(t)$ represents the moving average of confidence values over the current episode, $\sigma_{conf}(t)$ denotes the confidence stability measured as the standard deviation of recent confidence values, $\rho_{consec}(t)$ represents the normalized count of consecutive correct predictions, and $\psi_{stab}(t)$ measures the prediction stability between consecutive timesteps.

The action space $\mathcal{A} = \{0, 1\}$ provides binary decision options where action $a_t = 0$ indicates continuation of audio processing to incorporate additional frames, while $a_t = 1$ represents termination of processing with immediate classification output. The action selection is constrained by a minimum processing requirement: $a_t = 0$ if $t < t_{min}$ regardless of the agent's intended action, ensuring sufficient information accumulation before classification decisions.

The reward function $\mathcal{R}(s_t, a_t, s_{t+1})$ implements a sophisticated multi-modal optimization strategy that adapts to different operational requirements through parameterizable reward modes. The general formulation encompasses three distinct operational modes designed to optimize specific performance characteristics.

For the accuracy-first reward mode, prioritizing classification precision:
\begin{equation}
\mathcal{R}_{accuracy}(s_t, a_t, s_{t+1}) = \begin{cases}
\alpha_{acc} \cdot \mathbf{1}_{correct} + \beta_{conf} \cdot \mu_{conf}(t) + \mathcal{B}_{early}(t, \mu_{conf}) + \mathcal{B}_{consec}(t) - \sigma_{pen} \cdot (1-\mathbf{1}_{correct}) & \text{if } a_t = 1 \\
-\lambda_{time} + \mathcal{I}_{stable}(t) & \text{if } a_t = 0
\end{cases}
\end{equation}

For the balanced reward mode, optimizing both accuracy and efficiency equally:
\begin{equation}
\mathcal{R}_{balanced}(s_t, a_t, s_{t+1}) = \begin{cases}
\alpha_{acc}^{(b)} \cdot \mathbf{1}_{correct} + \beta_{conf}^{(b)} \cdot \mu_{conf}(t) + \mathcal{B}_{early}(t, \mu_{conf}) + \mathcal{B}_{consec}(t) - \sigma_{pen}^{(b)} \cdot (1-\mathbf{1}_{correct}) & \text{if } a_t = 1 \\
-\lambda_{time}^{(b)} & \text{if } a_t = 0
\end{cases}
\end{equation}

For the speed-focused mode, emphasizing computational efficiency:
\begin{equation}
\mathcal{R}_{speed}(s_t, a_t, s_{t+1}) = \begin{cases}
\alpha_{acc}^{(s)} \cdot \mathbf{1}_{correct} + \beta_{conf}^{(s)} \cdot \mu_{conf}(t) + \mathcal{B}_{early}(t, \mu_{conf}) + \mathcal{B}_{consec}(t) - \sigma_{pen}^{(s)} \cdot (1-\mathbf{1}_{correct}) & \text{if } a_t = 1 \\
-\lambda_{time}^{(s)} & \text{if } a_t = 0
\end{cases}
\end{equation}

The early stopping bonus function $\mathcal{B}_{early}(t, \mu_{conf})$ encourages efficient decision-making when confidence levels are sufficiently high and stable:

\begin{equation}
\mathcal{B}_{early}(t, \mu_{conf}) = \begin{cases}
\zeta_{bonus} \cdot \left(1 - \frac{t}{T_{max}}\right) & \text{if } t_{first\_correct} \leq t < t_{first\_correct} + 5 \text{ and } \mu_{conf}(t) > \tau_{threshold} \\
\zeta_{bonus} \cdot 0.5 \cdot \left(1 - \frac{t}{T_{max}}\right) & \text{otherwise if } \mu_{conf}(t) > \tau_{threshold} \\
0 & \text{otherwise}
\end{cases}
\end{equation}

The consecutive correct prediction bonus $\mathcal{B}_{consec}(t)$ rewards sustained accurate predictions:

\begin{equation}
\mathcal{B}_{consec}(t) = \begin{cases}
\xi_{consec} \cdot n_{consec} & \text{if } n_{consec} > 2 \\
0 & \text{otherwise}
\end{cases}
\end{equation}

where $n_{consec}$ is the number of consecutive correct predictions.

The stability incentive $\mathcal{I}_{stable}(t)$ for the continuation action is:

\begin{equation}
\mathcal{I}_{stable}(t) = \begin{cases}
-\omega_{stable} & \text{if } \mu_{conf}(t) > \tau_{threshold} \text{ and } \sigma_{conf}(t) < 0.02 \\
0 & \text{otherwise}
\end{cases}
\end{equation}

Through extensive experimental validation, we determined the optimal parameter configurations for each reward mode. For the accuracy-first mode, we set $\alpha_{acc} = 5.0$, $\beta_{conf} = 0.5$, $\sigma_{pen} = 5.0$, $\lambda_{time} = 0.0005$, and $\tau_{threshold} = 0.85$. The balanced mode employs $\alpha_{acc}^{(b)} = 3.0$, $\beta_{conf}^{(b)} = 0.3$, $\sigma_{pen}^{(b)} = 3.0$, and $\lambda_{time}^{(b)} = 0.001$. For speed-focused optimization, we utilize $\alpha_{acc}^{(s)} = 2.0$, $\beta_{conf}^{(s)} = 0.5$, $\sigma_{pen}^{(s)} = 2.0$, and $\lambda_{time}^{(s)} = 0.005$. Additional parameters include $\zeta_{bonus} = 0.2$, $\xi_{consec} = 0.1$, $\omega_{stable} = 0.01$, and $t_{min} = 15$.

Within this framework, the Deep Q-Network (DQN) with attention mechanism serves as the value function approximation and policy derivation mechanism for learning the optimal stopping strategy. The Q-function $Q_\theta(s_t, a_t)$ parameterized by neural network weights $\theta$ estimates the expected cumulative reward for taking action $a_t$ in state $s_t$, enabling the agent to learn context-dependent decision rules through interaction with the acoustic classification environment.

The DQN architecture incorporates a custom attention mechanism to enhance feature representation:

\begin{equation}
\mathbf{h}_t = f_{extract}(s_t; \theta_{extract})
\end{equation}

\begin{equation}
\mathbf{z}_t = \text{MultiHeadAttention}(\mathbf{h}_t, \mathbf{h}_t, \mathbf{h}_t; \theta_{attn})
\end{equation}

\begin{equation}
Q(s_t, a_t; \theta) = f_{head}(\mathbf{z}_t; \theta_{head})
\end{equation}
\noindent where $f_{extract}$ is the feature extraction network, MultiHeadAttention applies self-attention with 4 heads, and $f_{head}$ is the Q-value output head.

The DQN optimization objective minimizes the temporal difference error:

\begin{equation}
L(\theta) = \mathbb{E}_{(s,a,r,s') \sim \mathcal{D}}\left[\left(r + \gamma \max_{a'} Q_{\theta^-}(s', a') - Q_\theta(s, a)\right)^2\right]
\end{equation}

where $\mathcal{D}$ is the experience replay buffer, $\theta^-$ represents the target network parameters updated periodically with $\theta^- \leftarrow \tau \theta + (1-\tau)\theta^-$, and $\tau$ is the soft update coefficient.

The exploration strategy follows an $\epsilon$-greedy policy with linear decay:

\begin{equation}
\pi(a|s) = \begin{cases}
\arg\max_a Q_\theta(s, a) & \text{with probability } 1-\epsilon(t) \\
\text{uniform}(\mathcal{A}) & \text{with probability } \epsilon(t)
\end{cases}
\end{equation}

where $\epsilon(t) = \epsilon_{final} + (\epsilon_{initial} - \epsilon_{final}) \cdot \max(0, 1 - t/T_{explore})$.

The training process employs a multi-phase strategy to progressively balance accuracy and efficiency objectives, starting with accuracy-first mode and gradually transitioning to balanced optimization. This comprehensive framework enables adaptive optimization of the accuracy-efficiency trade-off through deep reinforcement learning methodologies tailored to acoustic-based vehicle speed classification requirements.

The application of reinforcement learning methodologies to acoustic processing tasks requires rigorous theoretical justification to ensure algorithmic validity and convergence guarantees. Reinforcement learning algorithms, particularly value-based methods like DQN, rely fundamentally on the assumption that the underlying system exhibits Markovian properties. Without establishing that our acoustic-based vehicle speed classification system satisfies the Markov property, the theoretical foundations for convergence to optimal policies, the validity of Bellman equations, and the applicability of dynamic programming principles would be compromised. The following is the proof of the Markov Property for Acoustic Processing Efficiency Optimization:

\textbf{Theorem 1 (Markov Property for Acoustic Processing Efficiency Optimization).} The acoustic-based vehicle speed classification system exhibits the Markov property, establishing the theoretical foundation for reinforcement learning applications in processing efficiency optimization.

\textbf{Proof.} We demonstrate that $P(S_{t+1} = s' | S_t = s, A_t = a, \mathcal{H}_t) = P(S_{t+1} = s' | S_t = s, A_t = a)$ where $\mathcal{H}_t$ represents the complete processing history, thereby validating the application of reinforcement learning methodologies for adaptive stopping strategies in acoustic classification tasks.

The state representation $s_t$ at timestep $t$ encapsulates all decision-relevant information for efficiency optimization through the comprehensive encoding as defined in Equation (10). The fundamental insight for acoustic processing efficiency lies in recognizing that the neural network's current predictions $\hat{y}_t = f_\theta(X_{1:t})$ represent a sufficient statistic for all accumulated acoustic evidence $X_{1:t} = \{x_1, x_2, \ldots, x_t\}$. The mapping function $f_\theta: \mathbb{R}^{t \times d_f} \rightarrow \mathbb{R}^C$ encapsulates the complete sequential processing of acoustic frames, where the confidence metrics and uncertainty measures derived from these predictions capture the essential characteristics needed for efficiency optimization decisions. 

The state transition follows the deterministic relationship:

\begin{equation}
s_{t+1} = \begin{cases}
g(s_t, x_{t+1}) & \text{if } a_t = 0 \text{ (continue processing)} \\
\text{terminal state} & \text{if } a_t = 1 \text{ (stop and classify)}
\end{cases}
\end{equation}

\noindent where $g: \mathcal{S} \times \mathbb{R}^{d_f} \rightarrow \mathcal{S}$ represents the deterministic state update function incorporating new acoustic information.

The temporal context completeness is established through the inclusion of normalized processing progress $\frac{t}{T_{max}}$, accumulated evidence quality metrics $\bar{\mu}_{conf}(t) = \frac{1}{k}\sum_{i=t-k+1}^t \mu_{conf}(i)$, and stability measures $\sigma_{conf}(t)$ and $\psi_{stab}(t)$ within the state representation. These components encode the current position within the allowable processing window and preserve temporal patterns relevant to stopping decisions without requiring explicit historical storage.

The reward structure for efficiency optimization exhibits locality properties through its dependence exclusively on current state information as defined in Equations (11)-(16), where accuracy assessment $\mathbf{1}_{correct}(\hat{y}_t)$, confidence bonuses $\mu_{conf}(t)$, and efficiency rewards are computed entirely from current state components.

To establish transition independence, we demonstrate that for any future state $s' \in \mathcal{S}$:

\begin{equation}
P(S_{t+1} = s' | S_t = s, A_t = a, \mathcal{H}_t) = P(S_{t+1} = s' | g(s, x_{t+1}), a) = P(S_{t+1} = s' | S_t = s, A_t = a)
\end{equation}

This equality holds because the state $s_t$ contains the sufficient statistics $\hat{y}_t$ and derived confidence measures that capture all decision-relevant information from the acoustic processing history. The transition probability depends only on the current neural network assessment encoded in $s_t$, the chosen action $a_t$, and the next acoustic frame $x_{t+1}$, which is independent of previous processing decisions.

Therefore, the acoustic processing efficiency optimization problem satisfies the Markov property, establishing that future processing states and optimal stopping decisions depend only on the current accumulated evidence and confidence metrics, independent of the specific sequence of previous processing steps. This theoretical foundation validates the application of reinforcement learning algorithms including value-based approaches and deep Q-learning architectures for learning adaptive stopping strategies that optimize the fundamental trade-off between classification accuracy and computational efficiency in acoustic-based vehicle speed classification systems. $\square$

\subsection{System Architecture}
\begin{figure*}[t]
	\centering
	\includegraphics[width=0.9\textwidth]{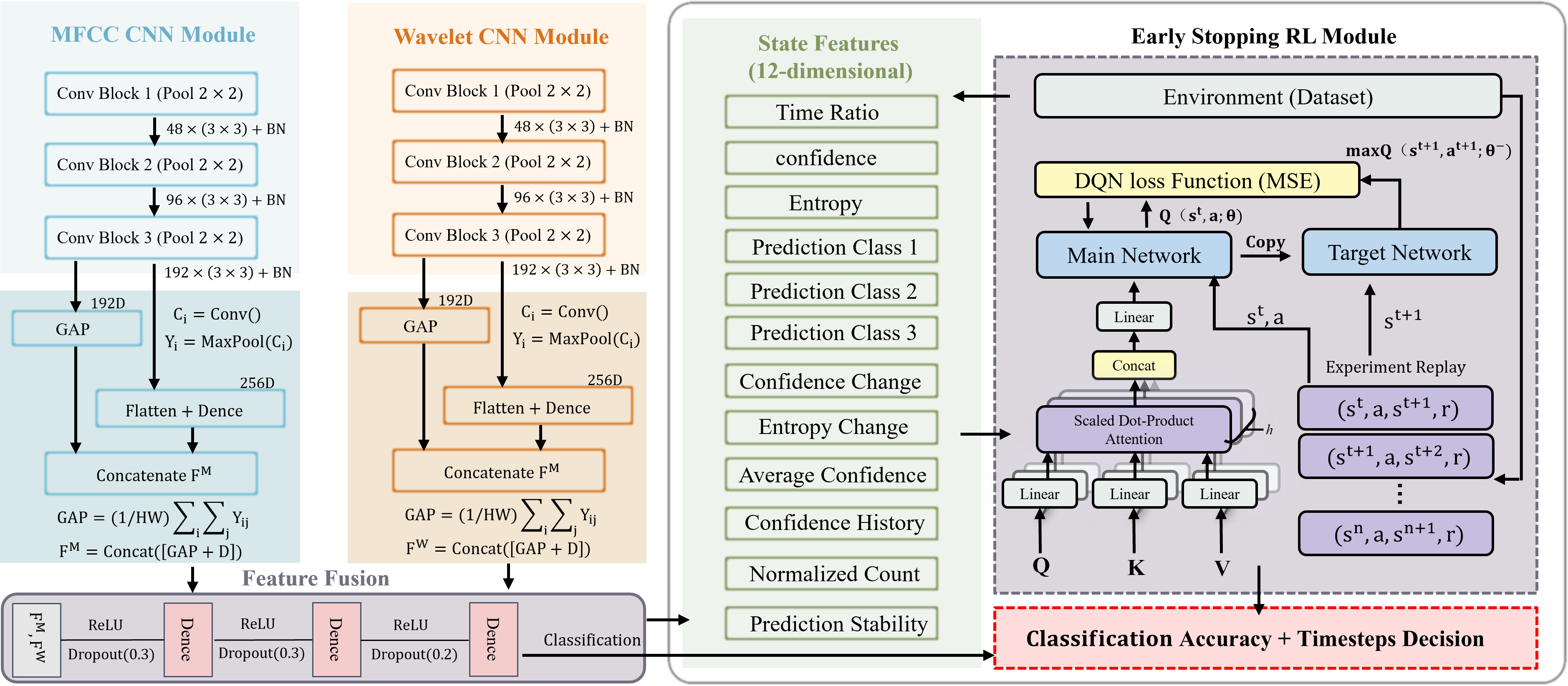}     
        \caption{Overall system architecture of the BMCNN-AttentionDQN framework.}
	\label{FS5}
\end{figure*}

The BMCNN-AttentionDQN algorithm integrates deep learning and reinforcement learning to achieve adaptive acoustic-based vehicle speed classification. Initially, the algorithm extracts MFCC and wavelet features from raw audio samples and trains a Bidirectional Multi-modal CNN (BMCNN) using cross-entropy loss. The BMCNN architecture processes both feature streams through separate CNN branches before fusion, establishing a robust base classifier. Subsequently, the trained BMCNN serves as a feature extractor within a reinforcement learning framework. The algorithm employs a Deep Q-Network with attention mechanism to learn an adaptive stopping policy. During training, the DQN processes partial audio sequences through the BMCNN, constructs a comprehensive state representation from the predictions, and decides whether to continue processing or classify immediately. The network is trained using experience replay and target network mechanisms to ensure stable learning. Through this approach, the system learns to adaptively determine the minimal number of frames required for confident classification, significantly reducing computational requirements while maintaining accuracy. The attention mechanism enhances the policy's ability to identify critical temporal patterns and confidence dynamics, leading to more informed stopping decisions. The BMCNN architecture is illustrated in Figure \ref{FS5}, and the corresponding pseudocode is presented in Algorithm 1.

\begin{algorithm}[t]
\SetAlgoLined
\caption{BMCNN-AttentionDQN Framework}
\label{alg:bmcnn_attentiondqn}
\KwIn{Audio dataset $\mathcal{D} = \{(\mathbf{x}_i, y_i)\}_{i=1}^N$}
\KwOut{Trained BMCNN $f_\theta^*$ and AttentionDQN policy $Q_\phi$}

Extract MFCC and Wavelet features for all samples in $\mathcal{D}$\\
Initialize BMCNN: $f_\theta \leftarrow$ BMCNN($\mathbf{X}_{\text{MFCC}}, \mathbf{X}_{\text{Wavelet}}$)\\
Train: $f_\theta^* \leftarrow \arg\min_\theta \mathcal{L}_{\text{CE}}(f_\theta(\mathbf{X}), y)$\\
Initialize: $Q_\phi$, $Q_{\phi^-}$, buffer $\mathcal{D} \leftarrow \emptyset$, $\epsilon \leftarrow \epsilon_{\text{initial}}$\\
\For{episode $= 1$ to $N_{\text{episodes}}$}{
    Sample $(\mathbf{X}_{\text{MFCC}}, \mathbf{X}_{\text{Wavelet}}, y)$; $t \leftarrow 1$\\
    \While{not done}{
        $\hat{\mathbf{y}}_t \leftarrow f_\theta^*(\mathbf{X}^{1:t})$; $s_t \leftarrow$ ConstructState($t, \hat{\mathbf{y}}_t$)\\
        $a_t \leftarrow \begin{cases}
            \text{Uniform}(\{0, 1\}) & \text{w.p. } \epsilon\\
            \arg\max_a Q_\phi(s_t, a) & \text{w.p. } 1-\epsilon
        \end{cases}$\\
        Execute $a_t$, get $r_t$; Store $(s_t, a_t, r_t, s_{t+1})$ in $\mathcal{D}$\\
        \If{$|\mathcal{D}| \geq$ batch\_size}{
            Update $Q_\phi$ using sampled minibatch from $\mathcal{D}$\\
        }
        \If{$a_t = 1$}{done $\leftarrow$ True}
        \Else{$t \leftarrow t + 1$}
    }
    Update: $\epsilon \leftarrow \max(\epsilon_{\text{final}}, \epsilon - \Delta\epsilon)$\\
    Periodically: $\phi^- \leftarrow \tau \phi + (1-\tau) \phi^-$
}
\Return $f_\theta^*$, $Q_\phi$
\end{algorithm}

\section{Experimental Setup and Implementation} \label{S5}
\subsection{Dataset Description and Preprocessing}
\begin{figure*}[h!]
	\centering
	\includegraphics[width=0.9\textwidth]{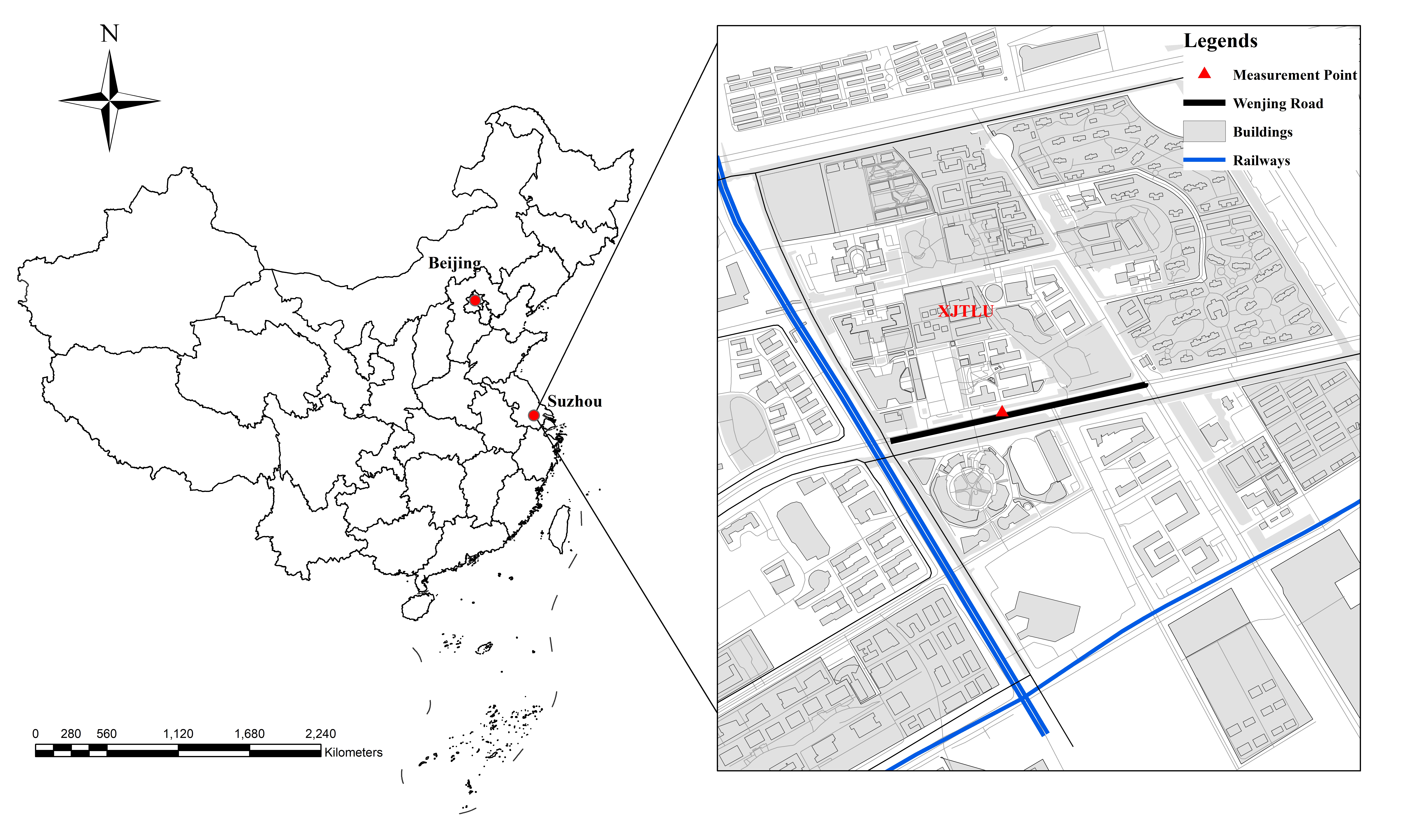}     
        \caption{Our measurement area.}
	\label{FS6}
\end{figure*}

The SZUR-Acoustic dataset was recorded on Wenjing Road, Suzhou (Jiangsu, China), a two-lane urban street separated from opposite traffic by a green belt as show in Figure \ref{FS6}. Using a smartphone placed 1.5 m above ground and a radar gun 50 m upstream, we captured 4 822 two-second audio clips: 339 samples labelled $30 km/h$, 3 215 samples labelled $50 km/h$ and 1 268 samples labelled $70 km/h$. This balanced yet realistically skewed distribution provides complementary low-, mid- and high-speed vehicle sounds for subsequent analysis.

The IDMT-Traffic corpus was captured at three German road sections that enforce speed limits of $30 km/h$ (Fraunhofer-IDMT), $50 km/h$ (Schleusinger-Allee) and $70 km/h$ (Langewiesener-Strasse). Ignoring vehicle type, every 2-s clip is assigned to one of these three speed classes. The dataset holds 17 086 recordings in total: 2 413, 9 145 and 5 528 samples for the 30, 50 and $70 km/h$ groups, respectively \cite{Dataset}.

\subsection{Baseline Model Configuration}
\begin{itemize}
\item \textbf{A3C (Asynchronous Advantage Actor-Critic)} \cite{based1}: 
A3C employs multiple parallel actors that asynchronously interact with the environment and update shared global parameters. The algorithm optimizes both policy and value functions simultaneously using policy gradient:
\begin{equation}
\nabla_\theta L_{\pi} = -\nabla_\theta \log \pi(a_t|s_t; \theta) \cdot A(s_t, a_t)
\end{equation}
and value function loss:
\begin{equation}
L_v = (R_t - V(s_t; \theta_v))^2
\end{equation}
where the advantage function is defined as:
\begin{equation}
A(s_t, a_t) = R_t - V(s_t)
\end{equation}
An entropy regularization term $\beta H(\pi)$ is added to encourage exploration.
\item \textbf{DDDQN (Dueling Double Deep Q-Network)} \cite{based2,based3}: 
DDDQN combines the Double DQN mechanism with a dueling network architecture to address overestimation bias and improve learning stability. The Q-function is decomposed as:
\begin{equation}
Q(s,a) = V(s) + \left(A(s,a) - \frac{1}{|\mathcal{A}|}\sum_{a'} A(s,a')\right)
\end{equation}
explicitly separating state value from action advantages. The target value is computed using:
\begin{equation}
y_t = r_t + \gamma Q(s_{t+1}, \arg\max_{a'} Q(s_{t+1}, a'; \theta); \theta^-)
\end{equation}
which decouples action selection from evaluation.
\item \textbf{SA2C (Soft Actor-Critic with Discrete Actions)} \cite{based4}: 
SA2C extends the maximum entropy reinforcement learning framework to discrete action spaces. The algorithm maximizes the objective:
\begin{equation}
J(\pi) = \mathbb{E}_{\rho_\pi} \left[\sum_t r(s_t, a_t) + \alpha \mathcal{H}(\pi(\cdot|s_t))\right]
\end{equation}
where $\alpha$ is an automatically adjusted temperature parameter that balances exploration and exploitation. This soft policy optimization approach maintains separate Q-networks and a policy network to achieve stable learning.
\item \textbf{PPO (Proximal Policy Optimization)} \cite{based5}: 
PPO constrains policy updates through a clipped surrogate objective to ensure stable learning. The algorithm optimizes:
\begin{equation}
L^{CLIP}(\theta) = \mathbb{E}_t\left[\min\left(r_t(\theta)\hat{A}_t, \text{clip}(r_t(\theta), 1-\epsilon, 1+\epsilon)\hat{A}_t\right)\right]
\end{equation}
where $r_t(\theta) = \frac{\pi_\theta(a_t|s_t)}{\pi_{\theta_{old}}(a_t|s_t)}$ is the probability ratio and $\epsilon$ is the clipping parameter. This approach prevents destructively large policy updates while maintaining sample efficiency. The value function is trained separately to minimize:
\begin{equation}
L^{VF}(\phi) = \mathbb{E}_t\left[(V_\phi(s_t) - V_t^{targ})^2\right]
\end{equation}
\item \textbf{TD3 (Twin Delayed Deep Deterministic Policy Gradient)} \cite{based6}: 
TD3 addresses function approximation errors in continuous control through three key techniques. It uses twin Q-networks with the target value:
\begin{equation}
y = r + \gamma \min_{i=1,2} Q_{\theta'_i}(s', \pi_{\phi'}(s') + \epsilon)
\end{equation}
to mitigate positive bias. Target policy smoothing adds noise:
\begin{equation}
\epsilon \sim \text{clip}(\mathcal{N}(0, \sigma), -c, c)
\end{equation}
to the target action. Additionally, the policy network is updated less frequently than the value networks (typically every $d=2$ iterations) to allow value estimates to stabilize.
\end{itemize}

\section{Result and Analysis}\label{S6}

This section presents a detailed account of the experimental results on the two acoustic datasets. The model's performance is thoroughly analyzed from multiple perspectives, including the impact of input sequence length, class-specific classification performance, training dynamics, and a comprehensive comparison against various baseline models.

\subsection{Influence of Input Sequence Length on Model Performance}

To investigate the impact of input data length on model performance, we conducted an ablation study by varying the sequence length, analyzed from two equivalent perspectives: frame utilization (as a percentage of the total duration) and absolute time window (in seconds). In this study, a full 2.00-second time window corresponds to 100\% frame utilization.

\begin{figure}[h!]
  \centering
  \begin{subfigure}[b]{0.49\linewidth}
    \includegraphics[width=\linewidth]{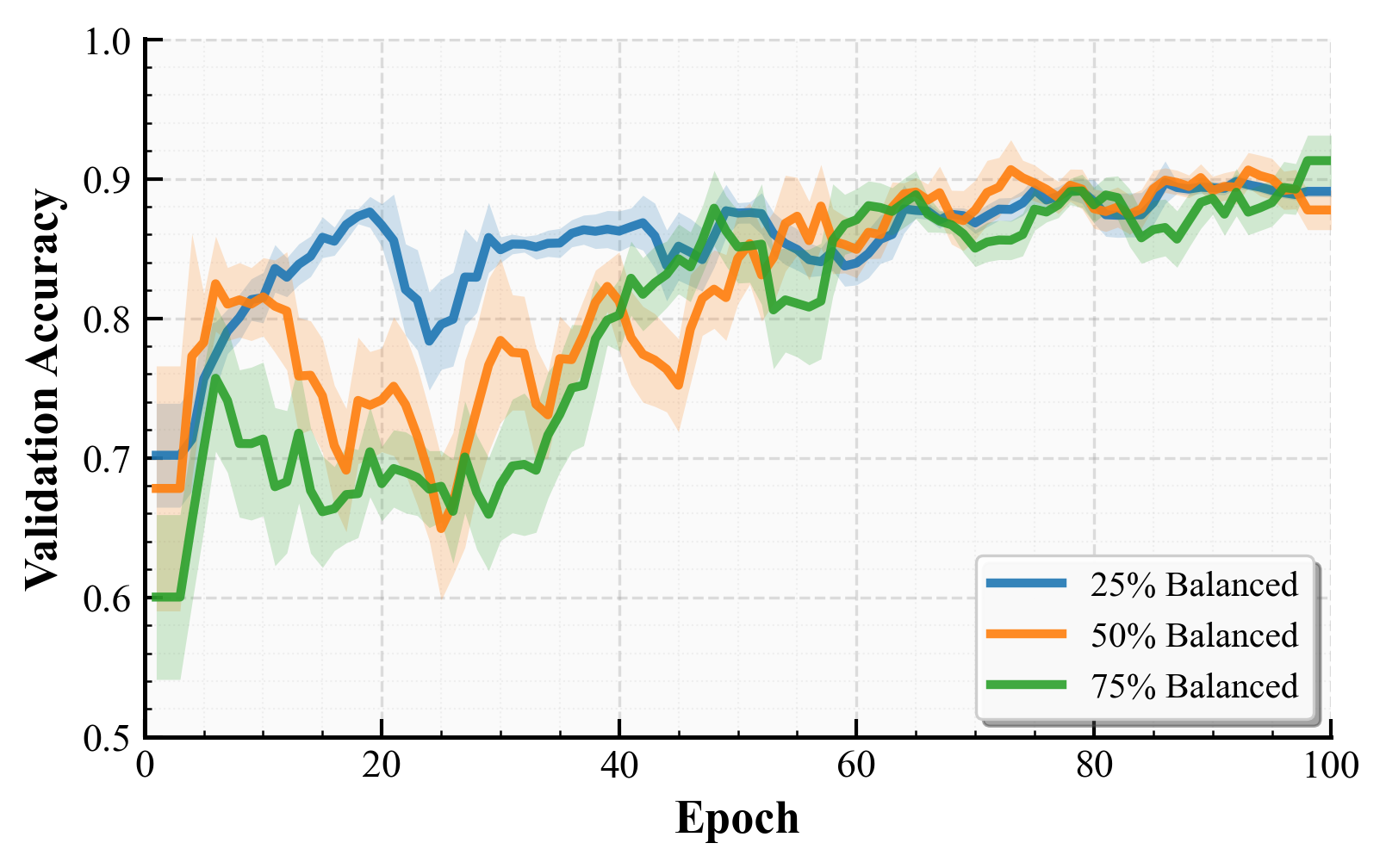}
    \caption{BMCNN (no RL) model with different frame utilization rates on IDMT Dataset}
  \end{subfigure}
  \hfill
  \begin{subfigure}[b]{0.49\linewidth}
    \includegraphics[width=\linewidth]{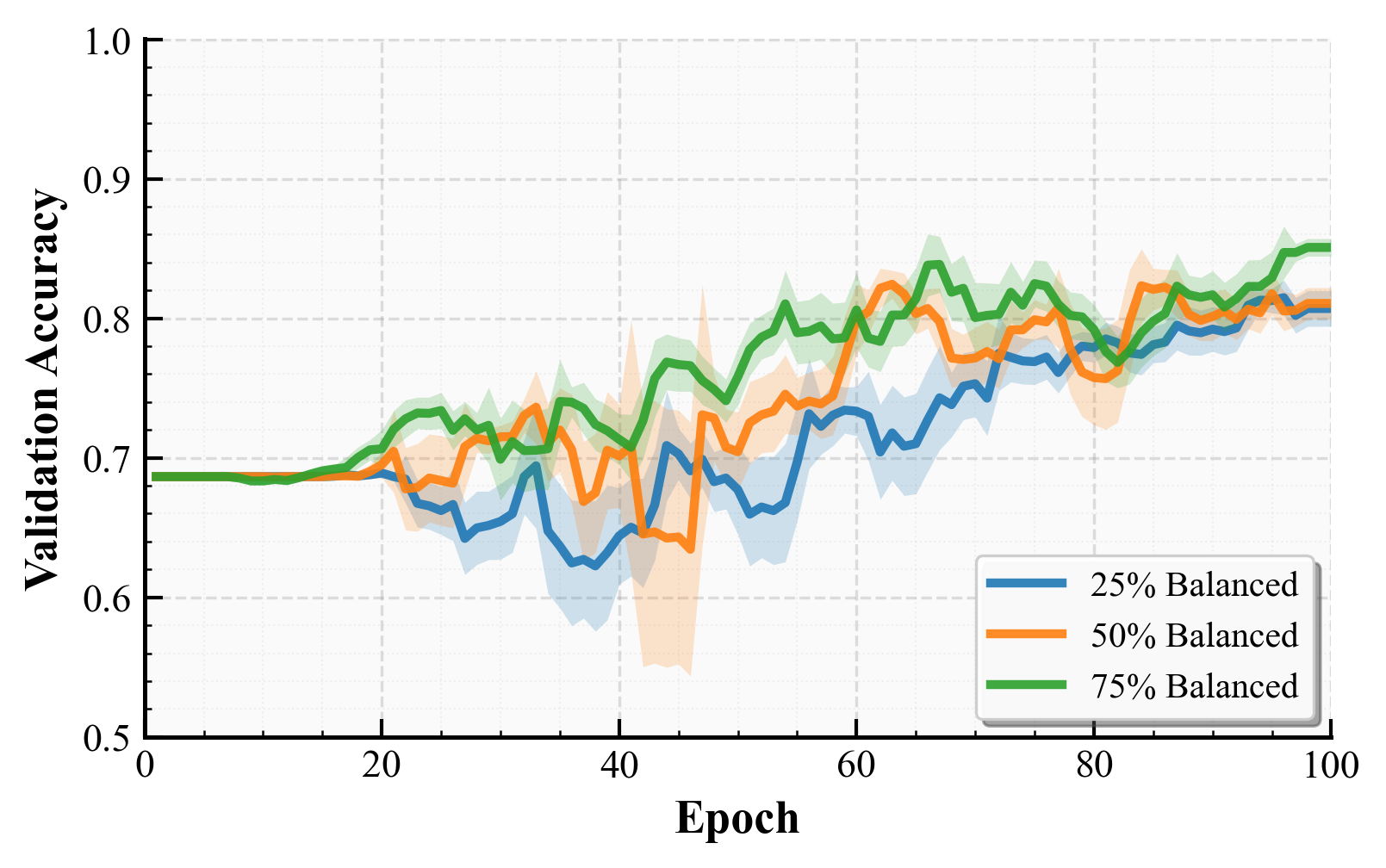}
    \caption{BMCNN (no RL) model with different frame utilization rates on SZUR-Acoustic Dataset}
  \end{subfigure}
  \vspace{1em}
  \begin{subfigure}[b]{0.49\linewidth}
    \includegraphics[width=\linewidth]{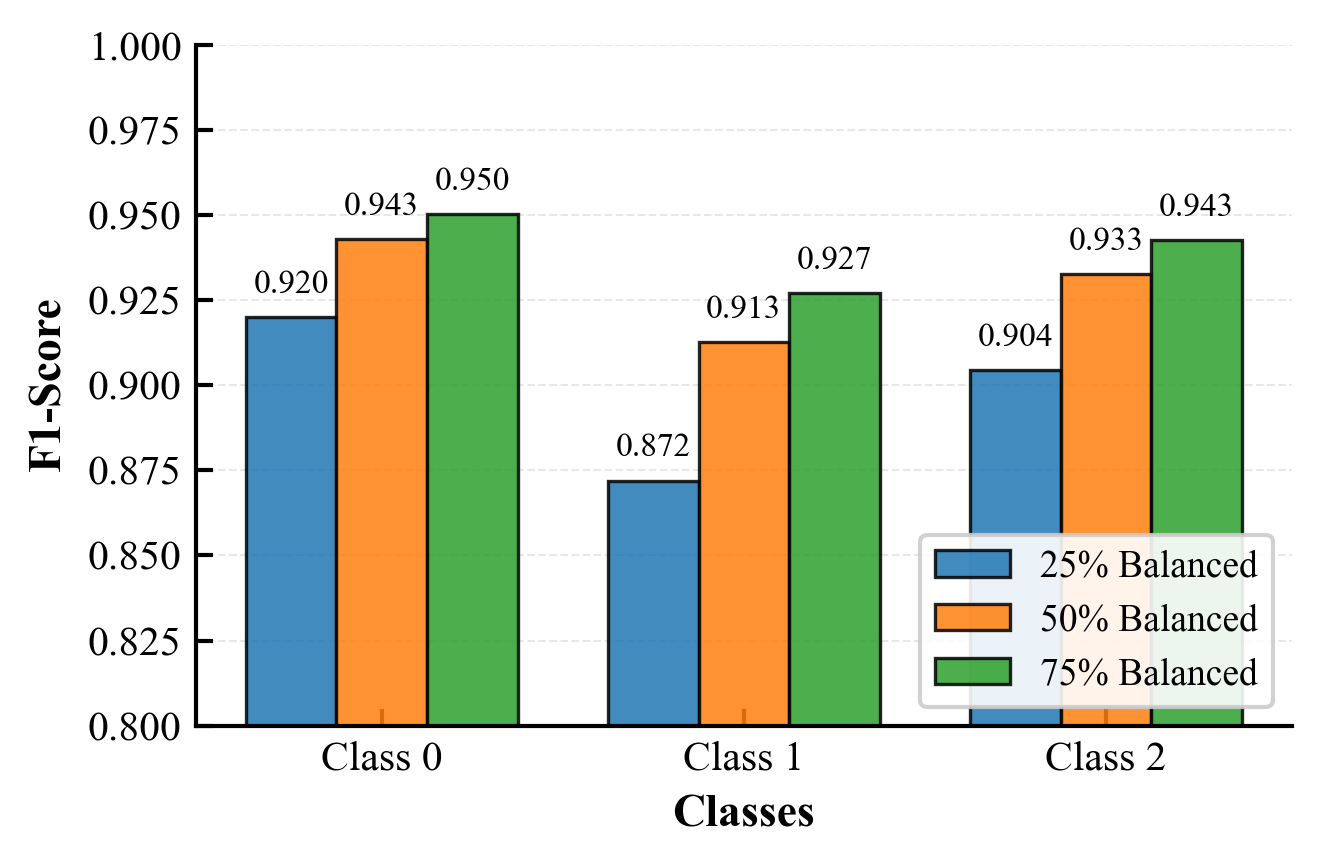}
    \caption{F1-Scores on IDMT Dataset}
  \end{subfigure}
  \hfill
  \begin{subfigure}[b]{0.49\linewidth}
    \includegraphics[width=\linewidth]{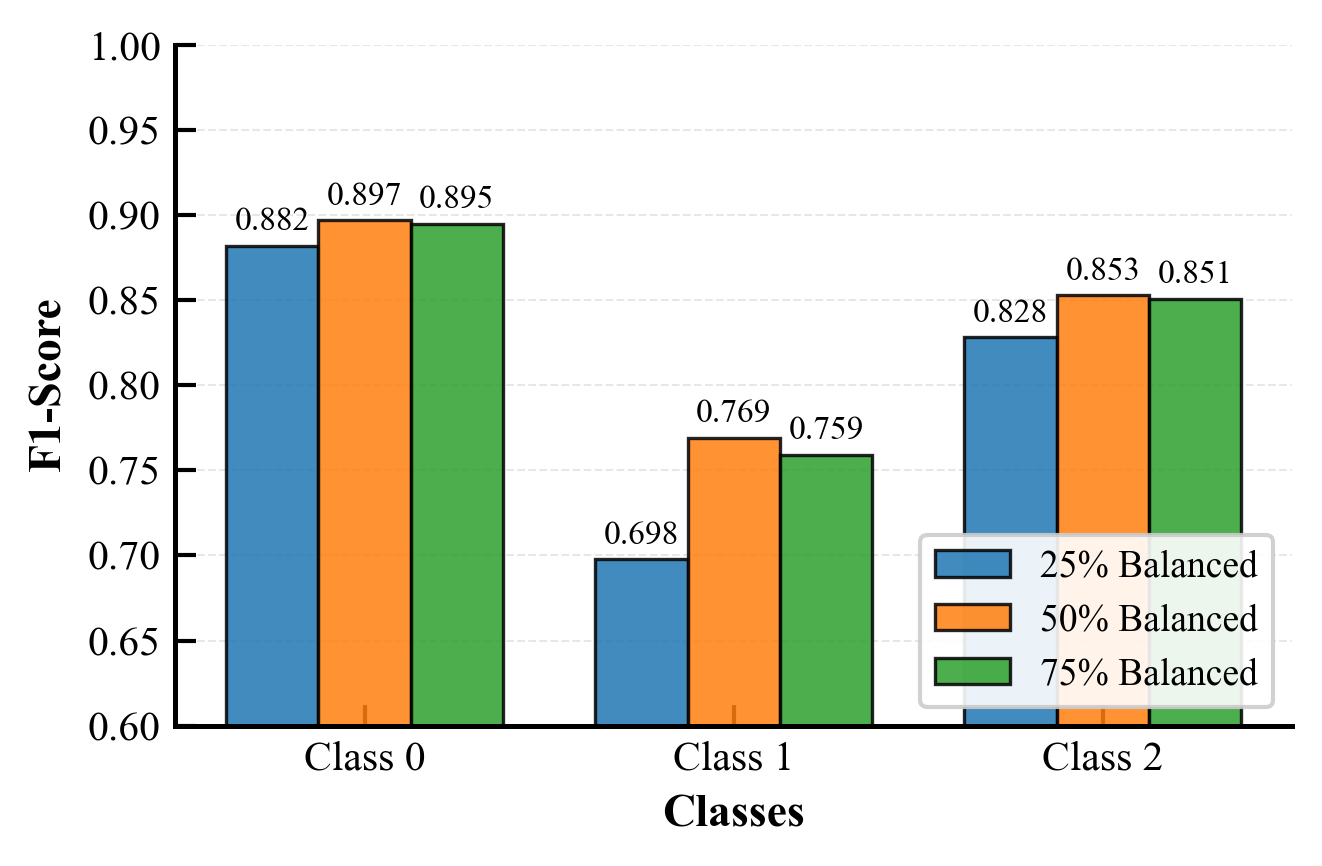}
    \caption{F1-Scores on SZUR-Acoustic Dataset}
  \end{subfigure}
  \caption{Impact of different frame utilization rates on BMCNN (no RL) model performance. (a, c) correspond to the IDMT dataset, and (b, d) correspond to the SZUR-Acoustic dataset.}
  \label{fig:frame_impact}
\end{figure}

As illustrated in Figure~\ref{fig:frame_impact}, increasing the input sequence length generally leads to a more stable and efficient convergence process. The training histories in Figures~\ref{fig:frame_impact}(a) and (b) show that increasing frame utilization from 25\% to 75\% improves both the validation accuracy and stability of the model. The F1-score comparison (Figures~\ref{fig:frame_impact}(c) and (d)) further quantifies this advantage, showing positive gains across nearly all classes on both datasets with higher utilization. This provides initial evidence that a longer temporal context helps the model capture more robust acoustic features.

Table~\ref{tab2}, which provides a performance breakdown at specific time windows, further validates and deepens these observations. The results clearly indicate that the \textbf{2.00-second time window (i.e., 100\% utilization) achieved the optimal overall accuracy on both datasets}. For the IDMT-Traffic dataset, the model's performance exhibited a \textbf{monotonic increase} with the time window length, reaching a peak accuracy of \textbf{95.99\%} at 2.00 seconds while achieving a balanced, high performance across all speed categories.

However, a more complex trade-off was observed on the SZUR-Acoustic dataset. Although the 2.00s window also secured the highest overall accuracy (\textbf{88.50\%}), its performance gain was primarily driven by a substantial advantage in the middle-speed category (\textbf{96.74\%}), which came at the cost of reduced performance in the low-speed (67.42\%) and high-speed (74.02\%) categories. Synthesizing the results from Figure~\ref{fig:frame_impact} and Table~\ref{tab2}, we can conclude that while a longer input sequence is generally more beneficial, the optimal length is not universal. It is closely tied to the intrinsic acoustic characteristics of the dataset and may involve a trade-off between overall accuracy and performance on specific subclasses.

\begin{table}[h!]
    \centering
    \caption{Performance of the BMCNN model (no RL) with different time.}
    \label{tab2}
    \begin{tabular}{cccccc}
        \toprule
        Dataset & Time (s) & Low Speed (\%) & Middle Speed (\%) & High Speed (\%) & Accuracy (\%) \\
        \midrule
        \multirow{4}{*}{IDMT-Traffic} & 0.25 & 92.70 & 88.02 & 94.42 & 90.43 \\
        & 0.50 & 96.07 & 90.13 & 98.53 & 93.27 \\
        & 1.50 & 96.96 & 91.41 & 98.77 & 94.27 \\
        & \textbf{2.00} & \textbf{95.03} & \textbf{98.14} & \textbf{92.86} & \textbf{95.99} \\
        \midrule
        \multirow{4}{*}{SZUR-Acoustic} & 0.25 & 80.43 & 82.37 & 85.23 & 82.80 \\
        & 0.50 & 72.73 & 84.90 & 90.43 & 85.28 \\
        & 1.50 & 70.77 & 85.15 & 89.78 & 85.08 \\
        & \textbf{2.00} & \textbf{67.42} & \textbf{96.74} & \textbf{74.02} & \textbf{88.50} \\
        \bottomrule
    \end{tabular}
\end{table}

\subsection{BMCNN-AttentionDQN Classification Performance Analysis}

\begin{figure}[h!]
  \centering
  \begin{subfigure}{0.49\linewidth}
    \includegraphics[width=\linewidth]{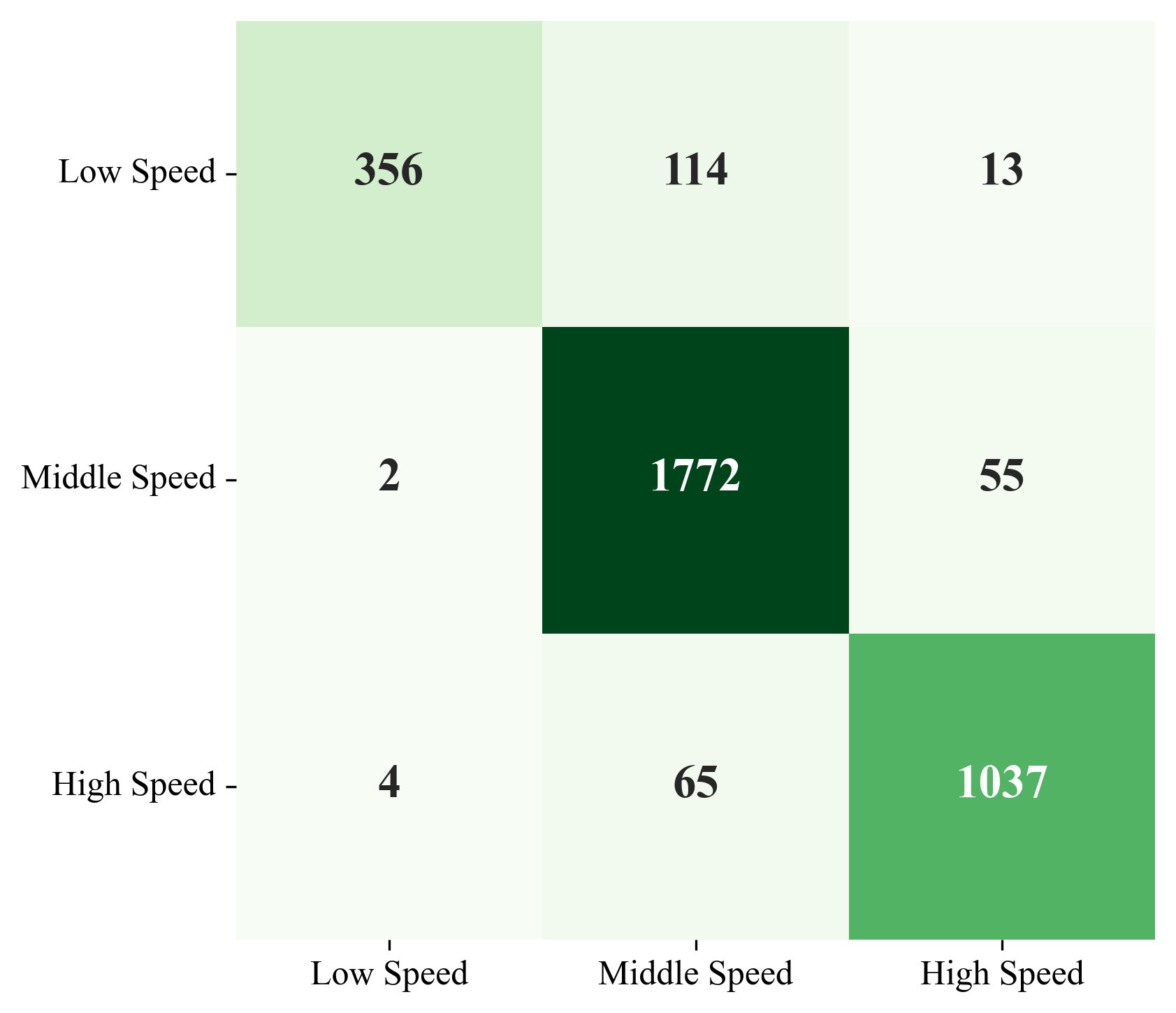}
    \caption{Confusion Matrix on IDMT Dataset}
    \label{fig:cm_a}
  \end{subfigure}
  \hfill
  \begin{subfigure}{0.49\linewidth}
    \includegraphics[width=\linewidth]{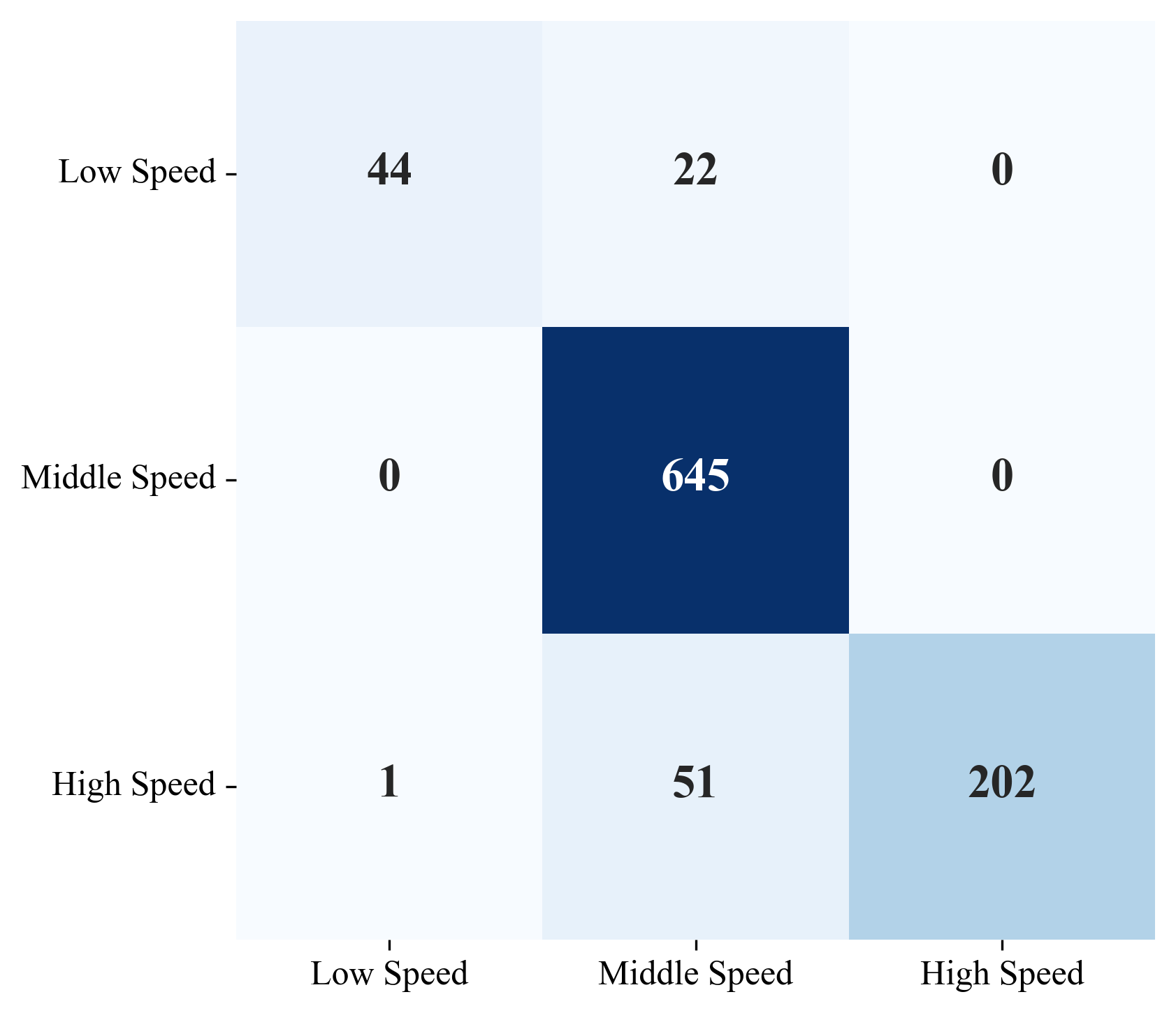}
    \caption{Confusion Matrix on SZUR-Acoustic Dataset}
    \label{fig:cm_b}
  \end{subfigure}
  \caption{Confusion matrices of the BMCNN-AttentionDQN on the (a) IDMT and (b) SZUR-Acoustic datasets.}
  \label{fig:confusion}
\end{figure}

To deeply investigate the class-level performance of the BMCNN-AttentionDQN, we generated confusion matrices for both datasets, as shown in Figure~\ref{fig:confusion}.

As depicted in Figure~\ref{fig:confusion}(a), the model demonstrated excellent classification capability on the IDMT dataset. The high values along the diagonal indicate that the vast majority of samples were correctly classified. In particular, \textbf{Class 1} achieved an exceptionally high \textbf{recall of 96.9\%} (1772 out of 1829). The primary source of confusion occurred between Class 0 and Class 1, with 114 samples from Class 0 being misclassified as Class 1, suggesting a similarity in their acoustic features.

On the more challenging SZUR-Acoustic dataset (Figure~\ref{fig:confusion}(b)), the model also performed robustly. \textbf{The recall for Class 1 reached a perfect 100\%} (645 out of 645), an outstanding result. However, the model encountered difficulty in distinguishing Class 0, achieving a \textbf{recall of only 66.7\%} (44 out of 66), with 22 samples being misclassified as Class 1. Furthermore, 51 samples from Class 2 were also misclassified as Class 1. This concentration of errors towards Class 1 suggests that its acoustic signature may be more generic or overlap with other classes, providing a clear direction for future optimization of the feature extraction network.

\subsection{BMCNN-AttentionDQN Training Dynamics}

\begin{figure}[h!]
  \centering
  \begin{subfigure}{0.49\linewidth}
    \includegraphics[width=\linewidth]{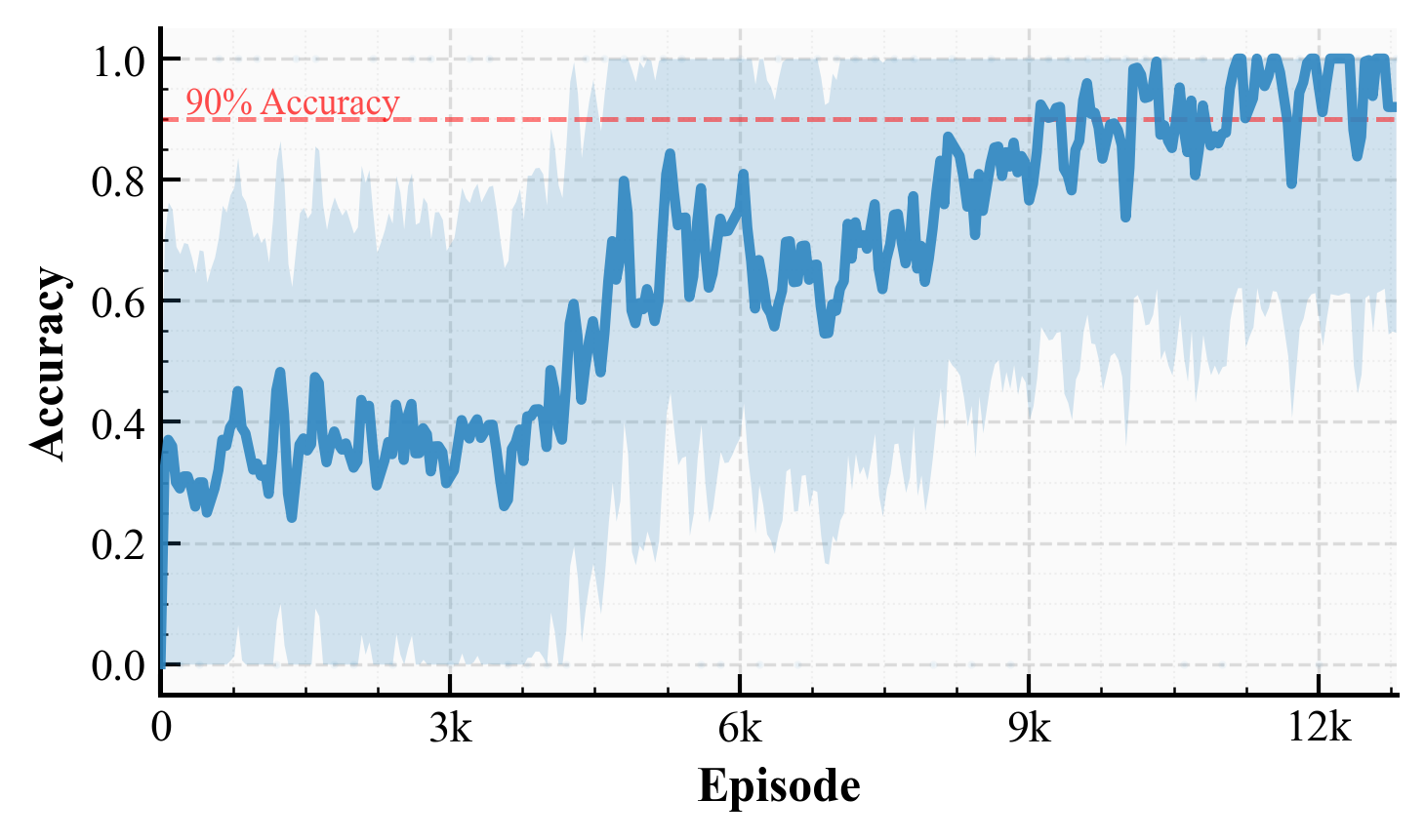}
    \caption{Training Process on IDMT Dataset}
    \label{fig:train_a}
  \end{subfigure}
  \hfill
  \begin{subfigure}{0.49\linewidth}
    \includegraphics[width=\linewidth]{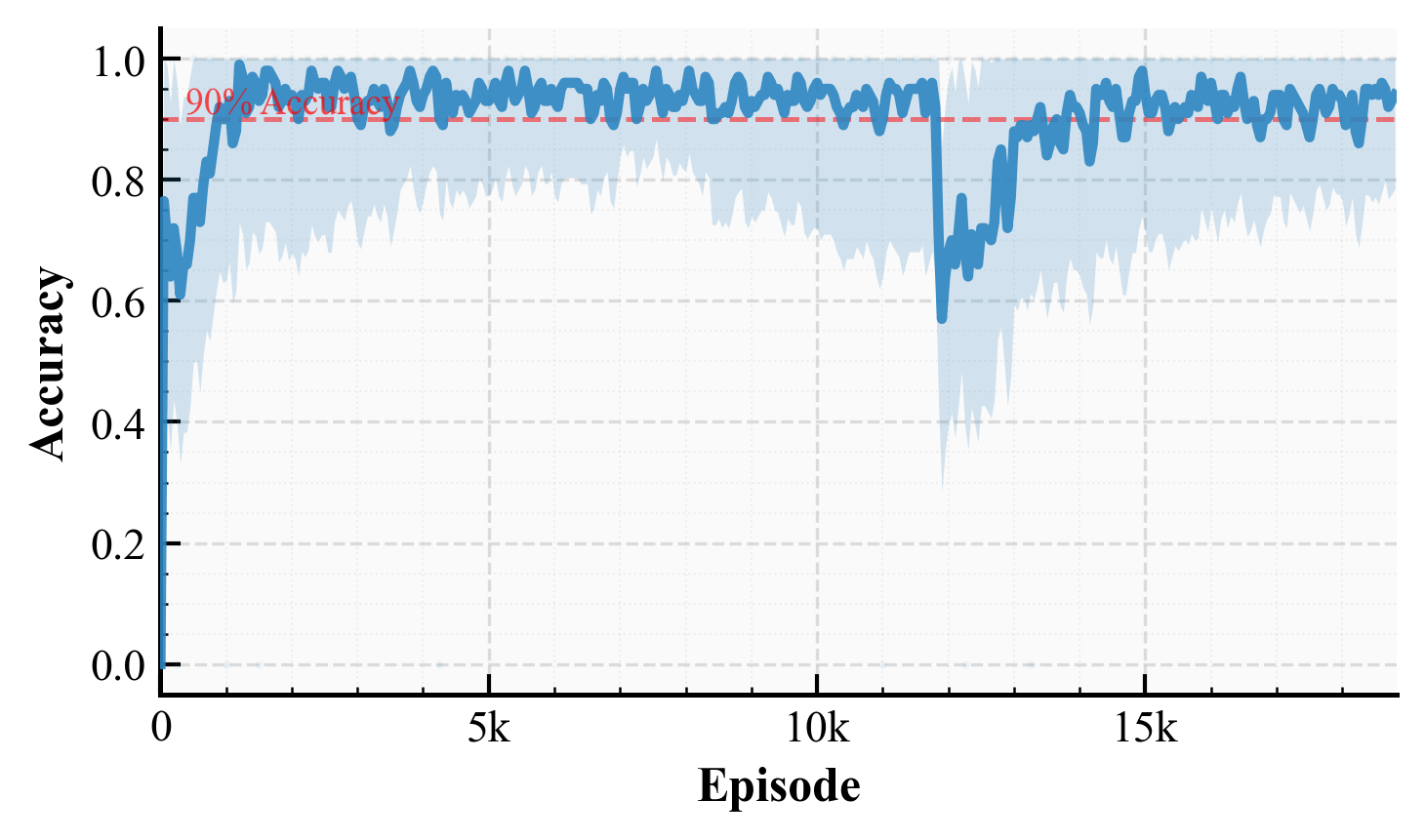}
    \caption{Training Process on SZUR-Acoustic Dataset}
    \label{fig:train_b}
  \end{subfigure}
  \caption{Training accuracy curves of the BMCNN-AttentionDQN on the (a) IDMT and (b) SZUR-Acoustic datasets.}
  \label{fig:training}
\end{figure}

The training process of the BMCNN-AttentionDQN, reflecting its learning efficiency and stability, is illustrated in Figure~\ref{fig:training}.

On the IDMT dataset (Figure~\ref{fig:training}(a)), the model exhibited \textbf{rapid and stable learning}. The accuracy quickly climbed above 80\% within approximately 6,000 episodes and ultimately converged to around 90\% after 10,000 episodes. The narrow confidence interval (shaded area) indicates high consistency and reproducibility across independent trials.

On the SZUR-Acoustic dataset (Figure~\ref{fig:training}(b)), the training process was more exploratory. The accuracy steadily improved from a lower baseline, surpassing 90\% after 12,000 episodes. Notably, a \textbf{transient dip in performance} occurred mid-training (around 12,000 episodes), which likely represents the DQN algorithm's exploration phase, escaping a local optimum to find a better policy. The model subsequently recovered swiftly and achieved a higher level of performance. Together, these training curves demonstrate that our DQN-Attention framework effectively balances exploration and exploitation, leading to stable convergence in complex acoustic fault detection tasks.

\subsection{Comprehensive Performance Comparison with Baseline Models}

\begin{figure}[h!]
  \centering
  \begin{subfigure}[b]{0.49\linewidth}
    \includegraphics[width=\linewidth]{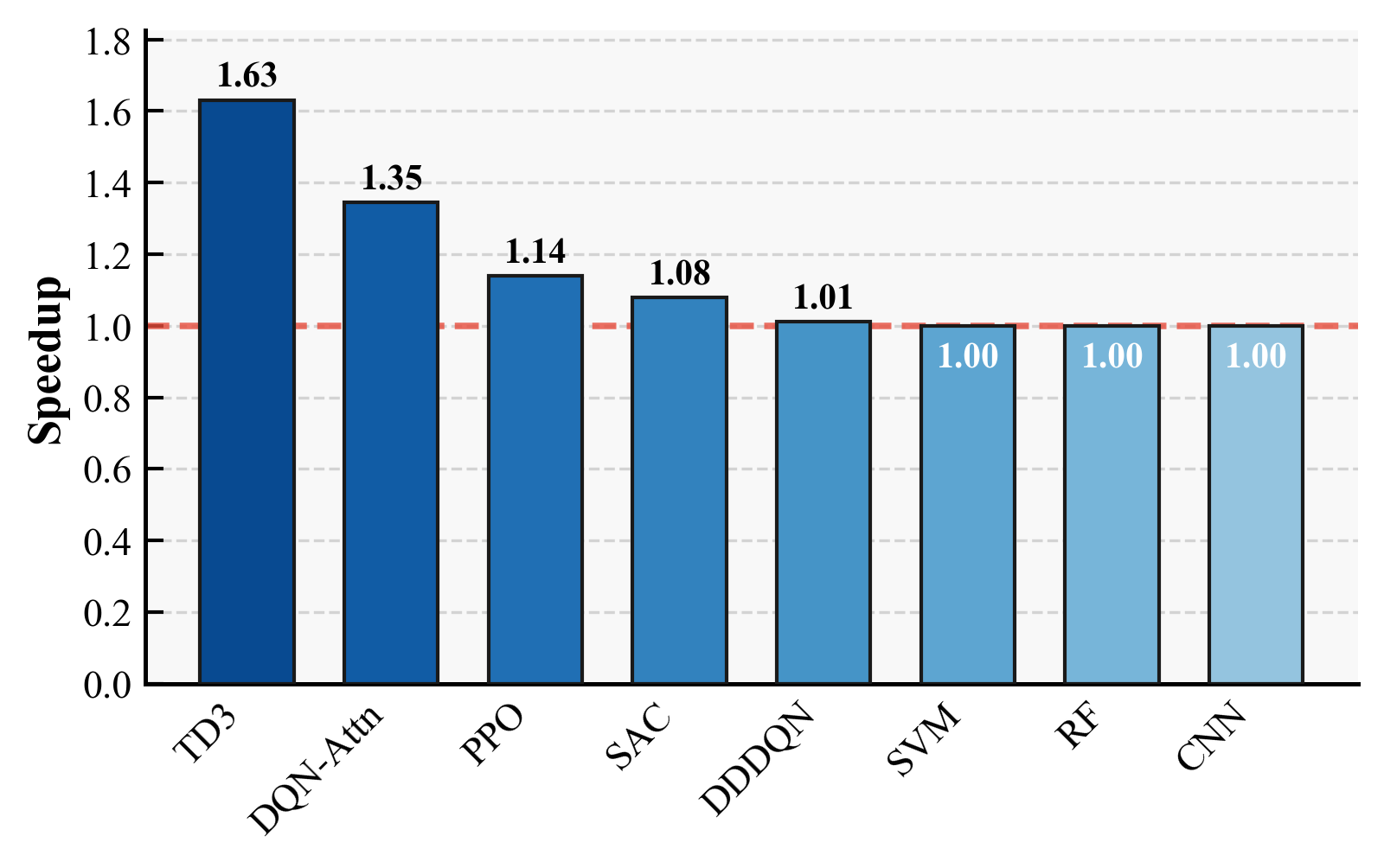}
    \caption{Speedup on IDMT Dataset}
    \label{fig:speed_a}
  \end{subfigure}
  \hfill
  \begin{subfigure}[b]{0.49\linewidth}
    \includegraphics[width=\linewidth]{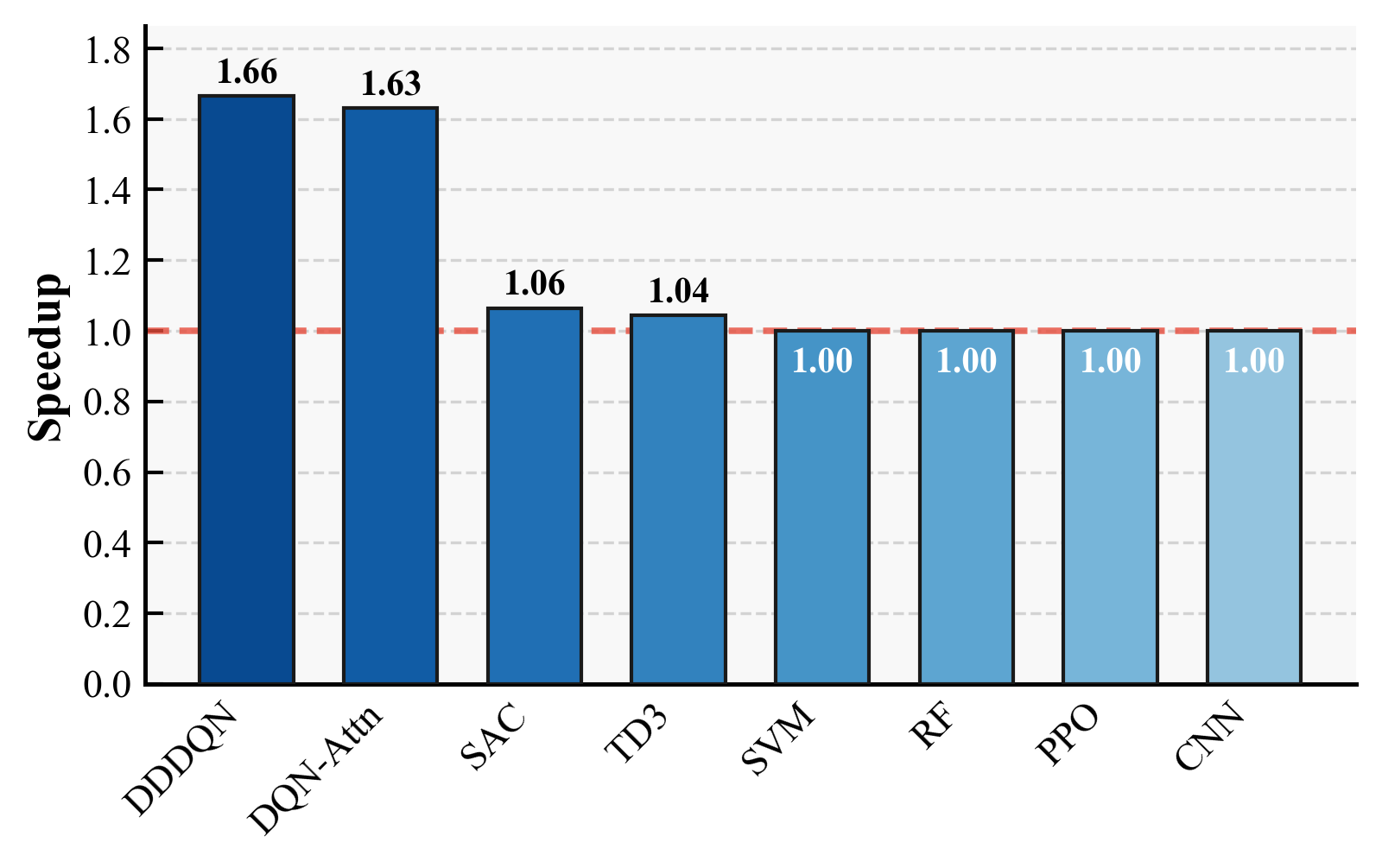}
    \caption{Speedup on SZUR-Acoustic Dataset}
    \label{fig:speed_b}
  \end{subfigure}
  \vspace{1em}
  \begin{subfigure}[b]{0.49\linewidth}
    \includegraphics[width=\linewidth]{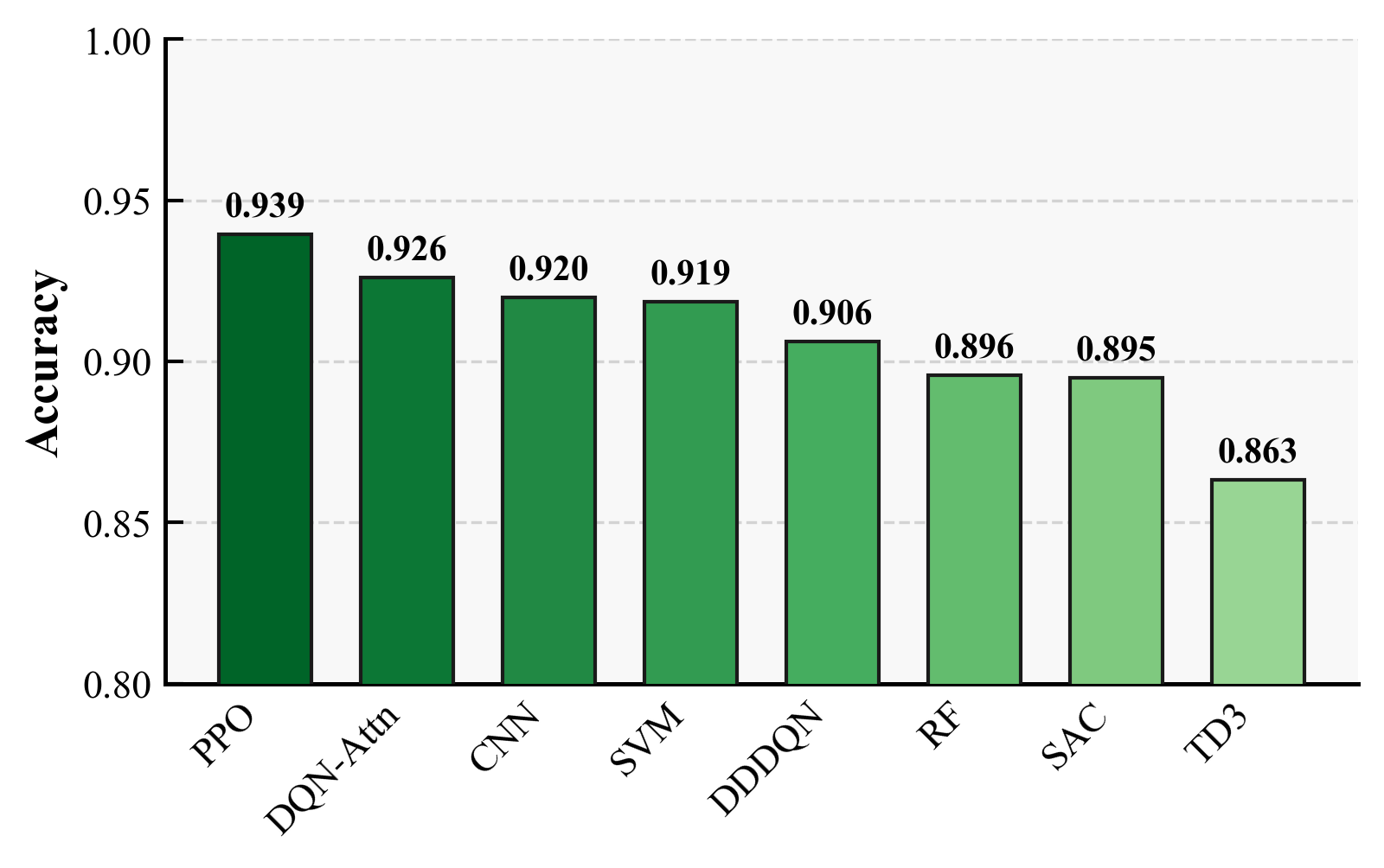}
    \caption{Accuracy on IDMT Dataset}
    \label{fig:acc_a}
  \end{subfigure}
  \hfill
  \begin{subfigure}[b]{0.49\linewidth}
    \includegraphics[width=\linewidth]{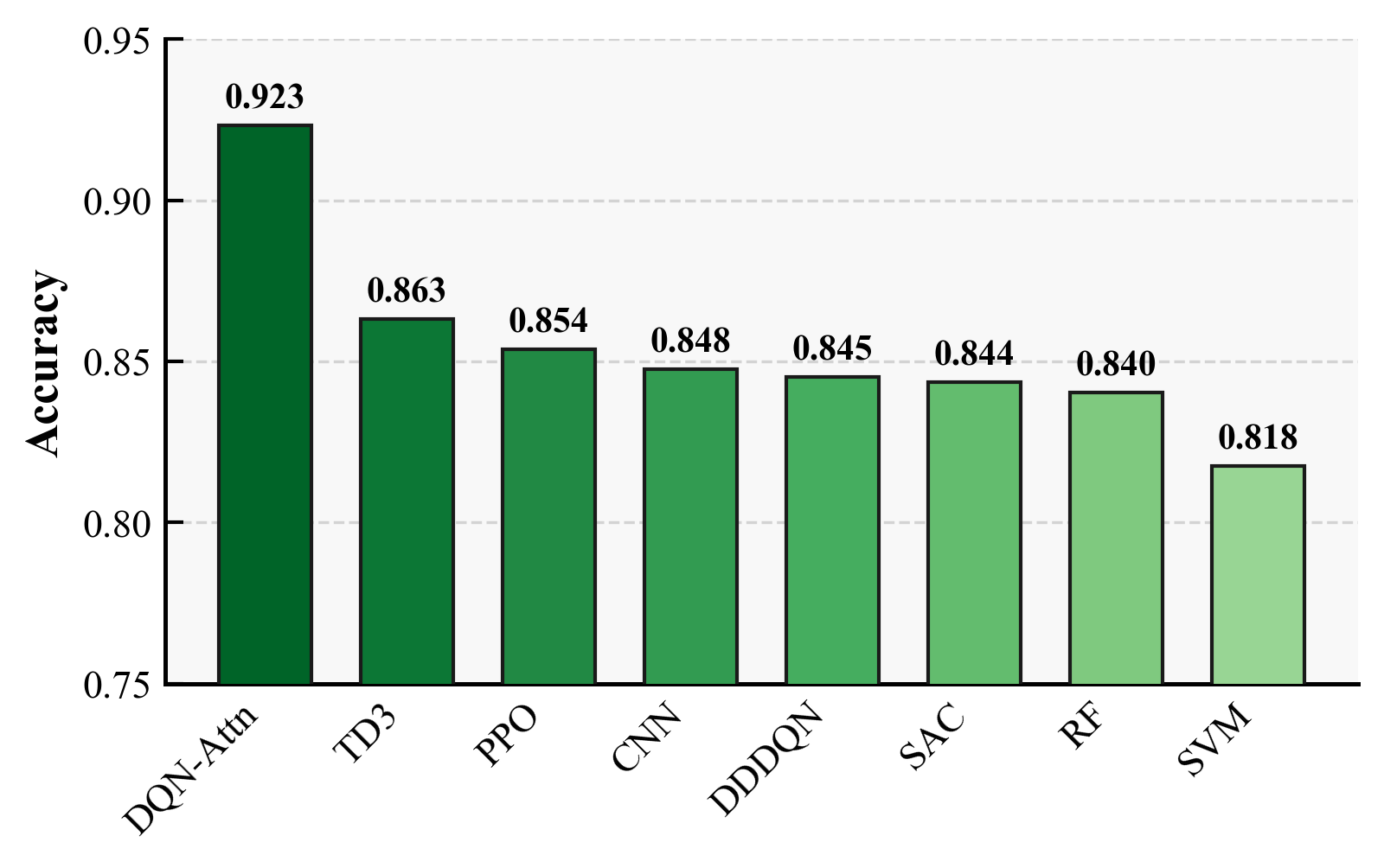}
    \caption{Accuracy on SZUR-Acoustic Dataset}
    \label{fig:acc_b}
  \end{subfigure}
  \caption{Comparison of speedup and accuracy for different models on the (a, c) IDMT and (b, d) SZUR-Acoustic datasets.}
  \label{fig:comparison}
\end{figure}

Finally, we conducted a comprehensive comparison of our BMCNN-AttentionDQN against several mainstream machine learning and deep learning baselines across two key dimensions: accuracy and computational efficiency (measured as speedup). The results are shown in Figure~\ref{fig:comparison}.

On the IDMT dataset (Figures~\ref{fig:comparison}(a) and (c)), the results reveal a \textbf{trade-off between accuracy and efficiency}. The \textbf{PPO} model achieved the highest accuracy (\textbf{0.939}) but with a modest speedup of 1.14x. In contrast, our \textbf{DQN-Attention} model ranked second in accuracy at \textbf{0.926} but delivered a superior speedup of \textbf{1.35x}, demonstrating a better overall balance.

On the SZUR-Acoustic dataset (Figures~\ref{fig:comparison}(b) and (d)), \textbf{the superiority of our BMCNN-AttentionDQN is unequivocal}. It not only achieved the highest accuracy at \textbf{0.923}, significantly outperforming all baselines (a 6-percentage-point margin over the runner-up, TD3, at 0.863), but it also attained a \textbf{1.63x speedup}, second only to DDDQN. This result strongly validates the \textbf{robustness and efficiency} of the BMCNN-AttentionDQN in handling complex acoustic data.

In summary, while other models may offer slightly higher speed (like TD3) or marginally better accuracy on a specific dataset (like PPO), our proposed DQN-Attention method consistently demonstrates \textbf{the best balance between accuracy and computational efficiency} across both datasets, positioning it as a highly competitive solution for real-world acoustic fault diagnosis applications.

\section{Conclusion and Future Work}\label{S7}

This paper presented a novel hybrid approach combining deep learning and reinforcement learning for acoustic-based vehicle speed classification in intelligent transportation systems. The proposed BMCNN-AttentionDQN framework successfully addresses the fundamental challenge of balancing classification accuracy with computational efficiency in real-time applications. Our key contributions include: (1) A Bidirectional Multi-modal Convolutional Neural Network (BMCNN) architecture that effectively processes dual acoustic representations (MFCC and wavelet features) through specialized CNN branches, achieving robust feature extraction from vehicle acoustic signatures; (2) An enhanced Deep Q-Network with attention mechanism that adaptively determines the optimal number of audio frames required for accurate classification, enabling intelligent early stopping decisions; and (3) Comprehensive experimental validation on both the IDMT-Traffic and SZUR-Acoustic datasets demonstrating superior performance compared to state-of-the-art baselines. The experimental results validate the effectiveness of our approach. On the IDMT dataset, the BMCNN-AttentionDQN achieved 92.6\% accuracy with a 1.35$\times$ speedup, demonstrating an optimal balance between performance metrics. On the more challenging SZUR-Acoustic dataset, our method achieved the highest accuracy of 92.3\% among all baselines while maintaining a 1.63$\times$ speedup. The attention mechanism proved particularly effective in identifying critical temporal patterns, enabling the system to make confident early classification decisions when appropriate.

Despite these promising results, several avenues for future research remain. Enhanced acoustic features beyond MFCC and wavelet representations, such as gammatone filterbanks or learned features through self-supervised learning, could further improve classification performance. The system requires evaluation across diverse acoustic environments including highways, tunnels, and varying weather conditions through domain adaptation techniques. Real-time hardware implementation on edge devices necessitates further optimization through model quantization, knowledge distillation, or specialized neural network accelerators. The framework could be extended beyond speed classification to simultaneous vehicle type identification, traffic density estimation, and anomaly detection, providing a comprehensive acoustic-based traffic monitoring solution. Additionally, integrating acoustic sensing with other modalities through sensor fusion could enhance system reliability while maintaining cost-effectiveness. The BMCNN-AttentionDQN framework represents a significant step toward practical acoustic-based traffic monitoring systems. By demonstrating that reinforcement learning can effectively optimize the accuracy-efficiency trade-off in real-world applications, this work opens new possibilities for cost-effective, scalable intelligent transportation solutions that can contribute to smarter, more efficient urban environments.

\clearpage

\bibliographystyle{elsarticle-num-names}

\bibliography{cas-refs}

\end{document}